\newcommand{\DEG}{\,^{\circ}}
\newcommand{\kms}{\,\textrm{km}/\textrm{s}}

\newcommand{\hinv}{\,h^{-1}}
\newcommand{\hkpc}{\,h^{-1}\,\textrm{kpc}}
\newcommand{\hmpc}{\,h^{-1}\,\textrm{Mpc}}
\newcommand{\mpc}{\textrm{Mpc}}
\newcommand{\kpc}{\textrm{kpc}}
\newcommand{\FD}{\textit{fracDeV}}
\newcommand{\DPA}{\Delta \textrm{PA}}
\newcommand{\DV}{\Delta V}
\newcommand{\DR}{\Delta R}
\newcommand{\lbol}{L_{bol}}

\newcommand{\SG}{\textrm{SG}}
\newcommand{\mSG}{\mathrm{SG}}
\newcommand{\HG}{\textrm{HG}}
\newcommand{\mHG}{\mathrm{HG}}
\newcommand{\isoa}{iso_A}
\newcommand{\isob}{iso_B}
\newcommand{\mmag}{\textit{modelMag}}
\newcommand{\reff}{R_{\mathit{eff}}}

\newcommand{\zconf}{\textit{zConf}}
\newcommand{\NSG}{N_{SG}}
\newcommand{\PAD}{\rm{PA}_{\rm{deV}}}
\newcommand{\PAI}{\rm{PA}_{\rm{iso}}}
\newcommand{\prat}{\mathcal{R}_{\mathrm{P}}}
\newcommand{\prad}{r_{\mathrm{p}}}

\newcommand{\qiso}{q_{iso}}
\newcommand{\qmom}{q_{mom}}
\newcommand{\hqiso}{\mathrm{HG}\;q_{iso}}
\newcommand{\hqmom}{\mathrm{HG}\;q_{mom}}
\newcommand{\sqiso}{\mathrm{SG}\;q_{iso}}
\newcommand{\sqmom}{\mathrm{SG}\;q_{mom}}

\documentclass[preprint,12pt]{aastex} 

\usepackage{graphics}
\usepackage{color}
\usepackage{ulem}
\usepackage{hyperref}

\shortauthors{Siverd, Ryden \& Gaudi}
\shorttitle{Satellite Galaxy Distribution and Alignment}

\begin{document}

\author{Robert Siverd}
\email{siverd@astronomy.ohio-state.edu}
\author{Barbara Ryden}
\email{ryden@astronomy.ohio-state.edu}
\author{B. Scott Gaudi}
\email{gaudi@astronomy.ohio-state.edu}
\affil{Department of Astronomy, The Ohio State University
    140 W. 18th Ave., Columbus, OH 43210}

\title{Galaxy Orientation and Alignment Effects in the SDSS DR6}


\begin{abstract} We identify, categorize, and quantify alignment effects among
host and satellite galaxies using a low-redshift ($z<0.23$) sample of
spectroscopically-confirmed galaxies from the Sloan Digital Sky Survey Data
Release 6. Consistent with other recent findings, we find that satellite
galaxies of red, centrally-concentrated (elliptical) host galaxies with radial
velocity separation $|\DV|<600\kms$ preferentially reside near the projected
major axes of their host galaxies. This preference is stronger among red,
centrally-concentrated satellite galaxies. We explore the dependence of this
satellite-host alignment on $\DV$ and the projected radial separation $\DR$,
finding that fractional anisotropy increases with decreasing $\DV$ and $\DR$.
Fractional anisotropy among the closest satellites ($\DR<250\hkpc$) is nearly
40\% greater than that seen among most distant ($500\hkpc<\DR<1000\hkpc$)
companions. We also investigate the effects of sample selection and measurement
errors on the measured alignment signals. Among highly concentrated satellite
galaxies at small projected separation ($\DR<300\hkpc$), we observe a strong
radial (hostward) alignment signal in isophotal position angles due to
isophotal twisting and contamination that is not present when using galaxy
model position angles. Among objects for which both isophotal and galaxy model
position angles agree to within $15\DEG$, this elongation signal is
significantly weaker. We also investigate the ``Holmberg Effect,'' a well-known
result wherein nearby ($<40\hkpc$) satellites of large, inclined spiral hosts
were seen to preferentially reside near the minor axes of their hosts in
projection. Due to the flux limit and spatial sampling limitations, a strict
test of the ``Holmberg Effect'' is not possible using only SDSS spectroscopic
galaxies. By adopting a looser set of cuts than those of Holmberg's original
study, we recover a comparable preference of faint blue satellites for the host
minor axis at marginal ($\sim3\sigma$) significance. After carefully inspecting
the methods used by various studies, we suggest that sample selection is
largely to blame for discrepant and contradictory results on galaxy alignment
effects in the literature. We conclude that several types of alignment likely
exist among different galaxy populations, but that the observed nature and
strength of those alignment trends depend sensitively not just on selection
criteria but also on the method used to determine galaxy orientation.
\end{abstract}


\section{Introduction} \label{s:intro}

The study of galaxy alignments is an old and contentious topic. The importance
of understanding galaxy alignment properties has grown over the past several
decades, as ever-increasing precision in numerical simulations has revealed
more about the galaxy formation process. The emergence of weak lensing as a
precision cosmology tool has further increased the demands on our understanding
of the intrinsic alignments of physically-related galaxies. An increase in the
number of conflicting results merits thorough re-examination of not only the
existing body of observational results, but also the methodologies used to
acquire those results, which may be responsible for these apparent
contradictions.


The angular distribution and alignment of satellite galaxies with respect to
their hosts carries information about the dynamical mechanisms involved in the
accretion and subsequent evolution of satellite galaxy halos. For example, it
is known from observational and theoretical studies that the distribution of
galaxies and their host halos is filamentary in nature. It has been suggested
that larger galaxies accrete satellites along these filaments
\citep{knebe_2004,libesk_2005,zentner_2005,bailin_2008}. Subsequent evolution
involves complex dynamical interactions between these satellites and the host
gravitational potential, including dynamical friction and tidal stripping.
Luminous galaxies within subhalos of larger cluster-size halos allow us to
trace the large-scale dark matter-dominated cluster potential. The signatures
of these dynamical processes may be imprinted on the alignments between the
central and satellite galaxies.

While a primary motivation for studying the intrinsic alignment of
physically-associated galaxies is to provide constraints for galaxy formation
models, understanding the magnitude of these alignments is important for
assessing possible contamination of weak lensing studies
\citep{croft_metzler_2000,lee_pen_2001}. Since weak lensing of background
galaxies by foreground substructures induces apparent alignments between these
lensed galaxies, intrinsic alignment of galaxies can provide a contaminating
signal to weak lensing studies. Calibrating the degree of this contamination
and its impact on precision cosmology with any given dataset requires an
accurate assessment of the magnitude and scale of intrinsic galaxy alignment
\citep[e.g.,][]{croft_metzler_2000,lee_pen_2001,bernberg_2002,mandelbaum_2005}.



Numerous alignment effects have been observed, many of which are only observed
in specific galaxy populations. One of the best known alignment effects, the
``Holmberg Effect'' \citep{holmberg_69}, is also currently one of the most
contentious. \citet{holmberg_69} noted that satellite galaxies (SGs)
preferentially occupy the space near the projected minor axis of very isolated,
large, and inclined spiral galaxies. After statistically correcting for
interloping field objects, there was a significant absence of satellites near
the host galaxy (HG) projected major axes. This would seem to indicate that
satellites preferentially travel or survive along polar orbits. This result has
been both rediscovered at marginal significance \citep[e.g.][]{zaritsky_97_ani}
and strongly refuted \citep[e.g.][]{brainerd_2005_ani,yang_2006,azzaro_2007} in
recent years. 


In the cases of refutation, the inverse effect has been observed. Specifically,
these studies found that satellites have a strong measured preference for the
major axis of their host galaxy as seen in projection. In both
\citet{yang_2006} and \citet{azzaro_2007}, this alignment exhibits a color
dependence, appearing strongest for satellites of red host galaxies. These red
galaxy populations are obviously dissimilar from those targeted by
\citet{holmberg_69}.


Galaxy clusters afford another opportunity for alignment analyses. For several
decades, focused observing campaigns and large-area surveys alike have
furnished evidence for anisotropy in the spatial distribution of cluster
members. \citet{Sastry_68}, using a sample of 9 Abell clusters, provided the
first evidence that cluster anisotropy might be common. Over the following
years, many more clusters were found to exhibit similar properties
\citep[e.g.][and many others]{austin_peach_74, dressler_78,
carter_metcalfe_1980, binggeli_1982}. Although conflicting results have
sporadically emerged \citep{tucker_peterson_1988,ulmer_1989}, it is generally
agreed that there exists a spatial anisotropy in favor of the projected major
axis of the brightest cluster galaxy (BCG). This effect is so strong that the
BCG major axis is a reliable indicator of the distribution of cluster
satellites \citep{binggeli_1982}. This is particularly useful when sample flux
limits may prevent detection of many cluster members. 

In addition to the alignment signal present within individual clusters,
alignment has been observed between neighboring clusters \citep[among
others]{binggeli_1982, ArgEtAl_1986, rhee_katgert_1987, west_1989a,
west_1989b}. Although lower in amplitude, this effect is observed to $10\,\mpc$
scales and beyond. Certain recent results have linked this alignment to the
local orientation of filamentary large-scale structures
\citep{bailin_2008,falten3}.


Several studies have also investigated the elongation of satellite galaxies
towards their hosts. Numerous studies \citep[e.g.,][]{croft_metzler_2000} have
suggested that an intrinsic alignment of this nature would mimic a weak lensing
signal and would require careful calibration. Other recent studies
\citep{pereira_kuhn_2005, AB_2006a, falten1} find significant evidence of
hostward elongation in an examination of galaxies from the Sloan Digital Sky
Survey \citep[SDSS;][]{DR4,DR3,SDSS_EDR,york2000} spectroscopic survey.
\citet{falten2} find a similar effect between dark matter halos in N-body
simulations. By contrast, \citet{bernberg_2002} observed no such elongation in
a study based on the 2 Degree Field Galaxy Redshift Survey
\citep[2dFGRS;][]{2dFGRS}.  


The size and nature of one's data set plays an important role in the detection
and study of galaxy alignment effects. Recent large-area surveys (e.g., 2dFGRS
\& SDSS) contain large numbers of galaxies, each with a uniform set of data
products. These greater numbers allow more stringent selection criteria while
still maintaining large statistical samples. The uniformity afforded by a fixed
reduction pipeline sidesteps most systematic problems with intercalibration of
different data sets. When available, spectroscopic redshifts can be used to
eliminate interlopers and greatly simplify selection of physically-related
galaxies.

On the other hand, smaller data sets often benefit from by-eye quality control
and selection which is impractical for very large surveys. Spurious results
that might go undetected in an automated pipeline can be found and removed.
Some smaller data campaigns also benefit from fainter limiting magnitudes,
which has proven important in studies of individual clusters.


The purpose of the present study is to independently investigate the alignment
effects mentioned above using the spectroscopic galaxy catalog of the Sloan
Digital Sky Survey Data Release 6 \citep[][hereafter SDSS DR6]{DR6}. We also
seek to resolve any discrepancies between our results and those that have been
presented to date.

Using a subsample of low-redshift ($z<0.23$) galaxies from the SDSS DR6
spectroscopic catalog, we explore the distribution and alignment properties of
satellite galaxies in projection. Groups are determined by centering a cylinder
of fixed physical size ($2400\kms$ in ``height,'' $1\hmpc$ in radius; Fig.\
\ref{fig:cylinder_angles}a) on the brightest local galaxy and counting objects
that fall within the cylinder volume. We impose a variety of different cuts in
observed galaxy properties and observe the resultant changes in satellite
galaxy distribution and alignment.

The main focus of this study is quality control in both selection criteria and
measurements of galaxy properties. To do this we simultaneously use multiple
independent measures of galaxy position angle and shape from SDSS. To determine
reliable relative luminosities, we incorporate all 5 photometric bands through
spectral templates, improving host determination among different galaxy types.
Lastly, we examine distribution and alignment properties separately as
functions of host and satellite color and concentration, host-satellite
velocity separation, and projected radial separation, combining and improving
upon results found in the literature.

The content of the present work is organized as follows. In \S\ref{s:data} we
describe the data products used in the study. In \S\ref{s:selection} we
describe the selection criteria used to pare the initial data set and detect
groups within it. In \S\ref{s:sample} we present the properties of our data
sample and discuss our methods of quality control. We present our findings in
\S\ref{s:results}. Numerous caveats and complications follow in
\S\ref{s:caveats}. How these results compare to previous findings follows in
\S\ref{s:comparison}. We present our investigation of the Holmberg Effect in
\S\ref{s:holmberg}. Lastly, a brief summary and discussion of our findings and
their implications can be found in \S\ref{s:discussion}.

In what follows we adopt the following cosmological parameters:
$\Omega_{\Lambda} = 0.7$, $\Omega_{M} = 0.3$, and $h \equiv H_0 / (100\,\kms /
\mpc) = 0.7$.


\section{Data Set} \label{s:data}

The galaxies we use are a subset of the SDSS DR6 \citep{DR6} spectroscopic
galaxy sample. In its entirety, the main spectroscopic catalog contains 790,220
galaxies to limiting Petrosian magnitude $m_r < 17.77$. We include all
spectroscopically-confirmed galaxies with well-established ($\zconf>0.35$)
redshifts of $0.004<z<0.23$. This lower bound safely excludes possible
neighbors of our own galaxy while the upper bound minimizes the impact of fiber
collisions. We further select galaxies with $g>0$ and $r>0$ (\mmag, see below).
Lastly we require successful (not placeholder) measurements of isophotal axes
($\isoa$ \& $\isob$). The resultant sample contains 572,495 galaxies which meet
these criteria.

SDSS employs matched de Vaucouleurs and exponential galaxy models to optimally
extract many photometric properties, including some that we use in this study.
A de Vaucouleurs $R^{1/4}$ profile accurately describes the light profiles of
many elliptical galaxies and spiral bulges. SDSS employs the following form
\citep{SDSS_EDR}:
\begin{equation} I(R) = I_0\;
\textrm{exp}\,[-7.67 \; ( R / \reff )^{1/4}\,].
\end{equation}
This profile is truncated beyond $7\reff$, decreasing smoothly to zero at
$8\reff$. The profile is softened slightly within $R<\reff / 50$.

Exponential light profiles are frequently used to describe dwarf ellipticals
and the disks of spiral galaxies. The SDSS exponential model, 
\begin{equation}
I(R) = I_0 \; \textrm{exp}\,[-1.68 \; ( R / \reff )], 
\end{equation} 
is truncated beyond $3\reff$, smoothly decreasing to zero at $4\reff$
\citep{SDSS_EDR}. 

The SDSS pipeline convolves two-dimensional variants of the above profiles with
a double-Gaussian approximation of the locally-determined PSF in order to
account for the effects of seeing. The best fit of the exponential and
deVaucouleurs models to each galaxy simultaneously determine the axis ratio,
position angle (PA, in degrees East from North), and effective scale radius
($\reff$, in arcseconds) in each case. The axis ratio and position angle are
taken to be constant throughout the model galaxy. Although fitted across a
range of radius, these models tend to best trace the bright, inner regions of
galaxies \citep{Strat01,SDSS_EDR}. 

SDSS provides several different galaxy flux measurements. Unbiased colors,
which are necessary to determine K-corrections to the highest possible
accuracy, must be obtained using the same aperture in each wavelength bands. To
accomplish this we employ $\mmag$ (model magnitude, hereafter simply magnitude)
fluxes in the present study. Both matched galaxy models (see above) are fitted
to each object. The fit of higher likelihood in $r$-band is re-applied to $u$,
$g$, $r$, $i$, and $z$, allowing only the amplitude to vary
\citep{SDSS_PHOTO,SDSS_EDR}. The resultant measurement, termed $\mmag$, thus
uses the same effective aperture in all bands. 

SDSS additionally computes a best-fit composite model flux in each band. This
\textit{cmodel} magnitude very accurately captures the total flux of each
galaxy by calculating the best-fit non-negative linear combination of
exponential and de Vaucouleurs profiles separately for each of the five
photometric bands. The fractional flux contribution of the de Vaucouleurs
profile, $\FD$, is a useful (seeing-corrected) measure of light profile
concentration.

The Petrosian flux, also computed for each galaxy, measures the light within a
circular aperture the radius of which is determined by the slope of the
azimuthally-averaged galaxy brightness profile. In the SDSS, this Petrosian
radius ($\prad$) is chosen such that $\prat(\prad) = 0.2$ \citep{SDSS_EDR},
where $\mathcal{R}_{\mathrm{P}}(r)$, the Petrosian ratio, is the ratio of
surface brightness in an annulus at radial position $r$ to the mean surface
brightness interior to $r$ \citep{blant01,yasuda01,SDSS_EDR}:
\begin{equation}
\prat(r) \equiv \frac{\int_{0.8r}^{1.25r} dr' \, 2\pi r' I(r') / [\pi(1.25^2 -
0.8^2) r^2]}{\int_0^r 2\pi r' I(r') / (\pi r^2)}.
\end{equation}
The Petrosian flux is then computed by summing the flux within $2r_p$. As
implemented by SDSS, $\prad$ measures the size of the bright inner regions of
galaxies. Its distribution closely matches (without artifacts) those of the
exponential and de Vaucouleurs model radii. Using $\prad$ we reject extremely
close neighbors that are likely heavily blended or misidentified bright knots
within a larger galaxy. In our study we exclusively use the $r$-band
measurement.

Galaxy light profile concentration is frequently measured using the inverse
concentration ratio ($C$), defined as the quotient of Petrosian 50\% and 90\%
radii [$C \equiv R_{50}/R_{90}$, where $R_{50}$ and $R_{90}$ are the radii that
contain $50\%$ and 90\% of Petrosian flux \citep{SDSS_EDR,blant01,yasuda01}].
$C$ is highly correlated with morphological type \citep{Shim01,Strat01} and
provides a convenient and effective way to statistically separate large
galaxies into different morphological classes. The Petrosian radii are not,
however, corrected for the effects of seeing, reducing the accuracy of
concentration ratio for smaller objects of sizes nearing that of the PSF.
Furthermore, as seeing is depending on the time of observation, relying on $C$
alone may introduce time-dependent systematic effects.

\begin{figure}
\epsscale{0.85}
\plotone{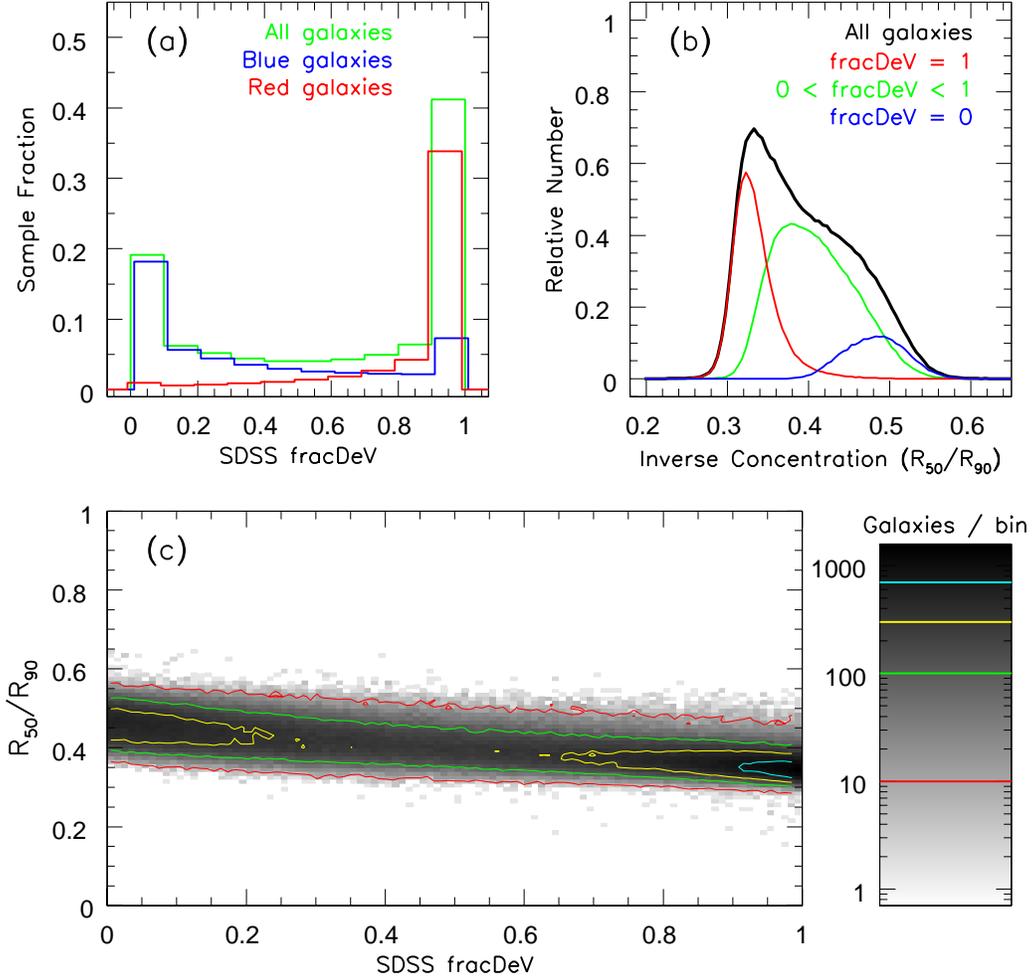} 
\epsscale{1.00}
\caption{\label{f:fracdev-color} Characteristics of SDSS galaxy parameter
$\FD$. Panel (a) shows the $\FD$ distribution of red and blue galaxies
separately as fractions of the total sample (572495 objects total). The red
and blue curves represent red and blue galaxies, respectively. The green curve
is the total of all galaxies. All galaxies in the initial sample are
represented. Panel (b) presents the full range of values for inverse
concentration (Petrosian $R_{50} / R_{90}$). Different values of $\FD$ are
isolated to illustrate the distinct groups that form the overall curve. The
sharp cutoff towards low inverse concentration is partially caused by the
effects of seeing. Panel (c) shows the strong correlation between fracDeV and
inverse concentration among galaxies of $0 \le \FD \le 1$. } 
\end{figure}

In the present work, we measure light profile concentration with SDSS galaxy
parameter \FD. Unlike Petrosian radii $R_{50}$ and $R_{90}$, the galaxy models
are corrected for seeing conditions (see above) and should thus provide a
straightforward and reliable measure of concentration which will depend less on
the angular scale of the individual galaxy relative to that of the PSF. We
illustrate the distributions of color and inverse concentration and how they
relate to $\FD$ in Figure \ref{f:fracdev-color}.

Although the galaxy fluxes are very well characterized, systematic effects in
the model-fitting procedure lead to some discretization in the determined model
scale radii and consequently into the model axis ratio \citep[][see
\S\ref{ss:discrepant_PA}]{DR6}. Therefore, to quantify the galaxy shape, we
instead calculate second-order adaptive moments \citep{bernjar,hirsel}. The
distribution of galaxy shapes determined in this way closely matches the
distribution determined by fitting the deVaucouleurs and exponential models,
but the adaptive moment method is free of the model-fitting artifacts. 

Adaptive moments shape measures are easily calculated using data provided by
SDSS. SDSS DR6 uses a Gaussian weight function $w(x,y)$ matched to the shape
and size of the galaxy image $I(x,y)$. The first-order moments
\begin{equation}
\mathbf{x_0} = \frac{\int \mathbf{x}\,w(x,y)\,I(x,y)\,dx\,dy}{\int
w(x,y)\,I(x,y)\,dx\,dy}
\end{equation}
yield the galaxy positions. Using this value, the second-order adaptive moments
can be computed as in
\begin{equation}
M_{xx} = \frac{\int (x-x_0)^2 \, w(x,y) \, I(x,y) \, dx \, dy}{\int w(x,y) \,
I(x,y) \, dx \, dy}.
\end{equation}
The SDSS DR6 provides the second-order parameters $\tau$, $e_{+}$ and
$e_{\times}$ for all galaxies, where $\tau = M_{xx} + M_{yy}$, $e_{+} = (M_{xx}
- M_{yy})/\tau$, and $e_{\times} = 2M_{xy}/\tau$. Using these values, we define
the adaptive moments axis ratio
\begin{equation}
\qmom = \left ( \frac{1 - e}{1 + e} \right )^{1/2},
\end{equation}
where $e = (e_{+}^2 + e_{\times}^2)^{1/2}$ \citep{ryden_04}. This shape is not
corrected for effects of seeing. For objects larger than the PSF (true of the
objects in our sample), however, any such correction will be negligible. In our
analysis we exclusively adopt the value of $\qmom$ determined from the $r$-band
measurements.  

SDSS provides model-independent measurements of PA and ellipticity by fitting
to the 25 magnitudes / arcsec$^2$ isophote. After first determining the object
centroid, the reduction pipeline determines the radius of this isophote as a
function of angle on the sky. This angular profile is then expanded in a
Fourier series, the coefficients of which determine the isophotal major
($\isoa$) and minor ($\isob$) axes, plus position angle. Although this
procedure is performed separately in all 5 photometric bands, we use the
$r$-band PA exclusively for both isophotal and galaxy model PA.

Although anticipated in a future Data Release, SDSS does not presently provide
error estimates for its position angle (PA) measurements. We note, however,
that the galaxies in the spectroscopic sample are considerably brighter than
the SDSS photometric limit. As a result, for those objects that meet our shape
criteria (see \S\ref{ss:angles}), PA measurement errors should
be negligible. We do, however, observe significant systematic errors in the
PA measurements. These must therefore be treated with care (see
\S\ref{ss:discrepant_PA}).

From $\isoa$ and $\isob$, we compute both the axis ratio and an isophotal
``radius.'' We define isophotal axis ratio $\qiso \equiv \isoa / \isob$ ($0 <
\qiso \le 1$) and isophotal radius $iso_R \equiv \sqrt{\isoa\,\isob}$. We use
the latter to exclude spurious satellite galaxies (e.g., blended objects or
bright substructure) and to eliminate possible cases of isophotal contamination
(see \S\ref{ss:discrepant_PA}).

In practice, the 25 mag arcsec$^{-2}$ isophote tends to trace the outermost
observable regions of galaxies. The measured isophotal PA may, in fact, be very
different from the PA determined from fitting models to the galaxy light
distribution. In our analysis, we compute the position angle difference
($\DPA$) between isophotal and model PAs which we then use to identify cases
where one (or both) PA measurements are unreliable or where the two PA
measurements are genuinely discrepant. Several recent studies
\citep{brainerd_2005_ani,yang_2006,falten1} have reported alignment findings
based on isophotal PA alone. We believe that these two different PA
measurements can have different physical meaning with respect to galaxy
structure. As a result, both should be considered. We explore the effects of
using one or the other PA determination in \S\ref{ss:discrepant_PA}.

Following \citet{bailin_2008}, we K-correct our galaxy colors to $z = 0.1$.
Selection of this ``neutral'' redshift ($z = 0.1$ is both the mean and median
of redshift our sample) minimizes K-correction error for the bulk of our
galaxies. We denote K-corrected magnitudes in the usual way with the redshift
in superscript (e.g., $^{0.1}r$). To perform the corrections, we employ the
Low-Resolution Templates\footnote{The Low-Resolution Templates of
\citet{roberto} are a collection of publicly available Fortran routines. See
\url{http://www.astronomy.ohio-state.edu/~rjassef/lrt} for more information.}
(LRT) software package of \citet{roberto}. This package also uses these
templates to calculate the bolometric luminosity (where $\lbol$ spans 0.2 to 10
$\mu m$), which we use to robustly separate host and satellite galaxies. 

\begin{figure} 
%
\plotone{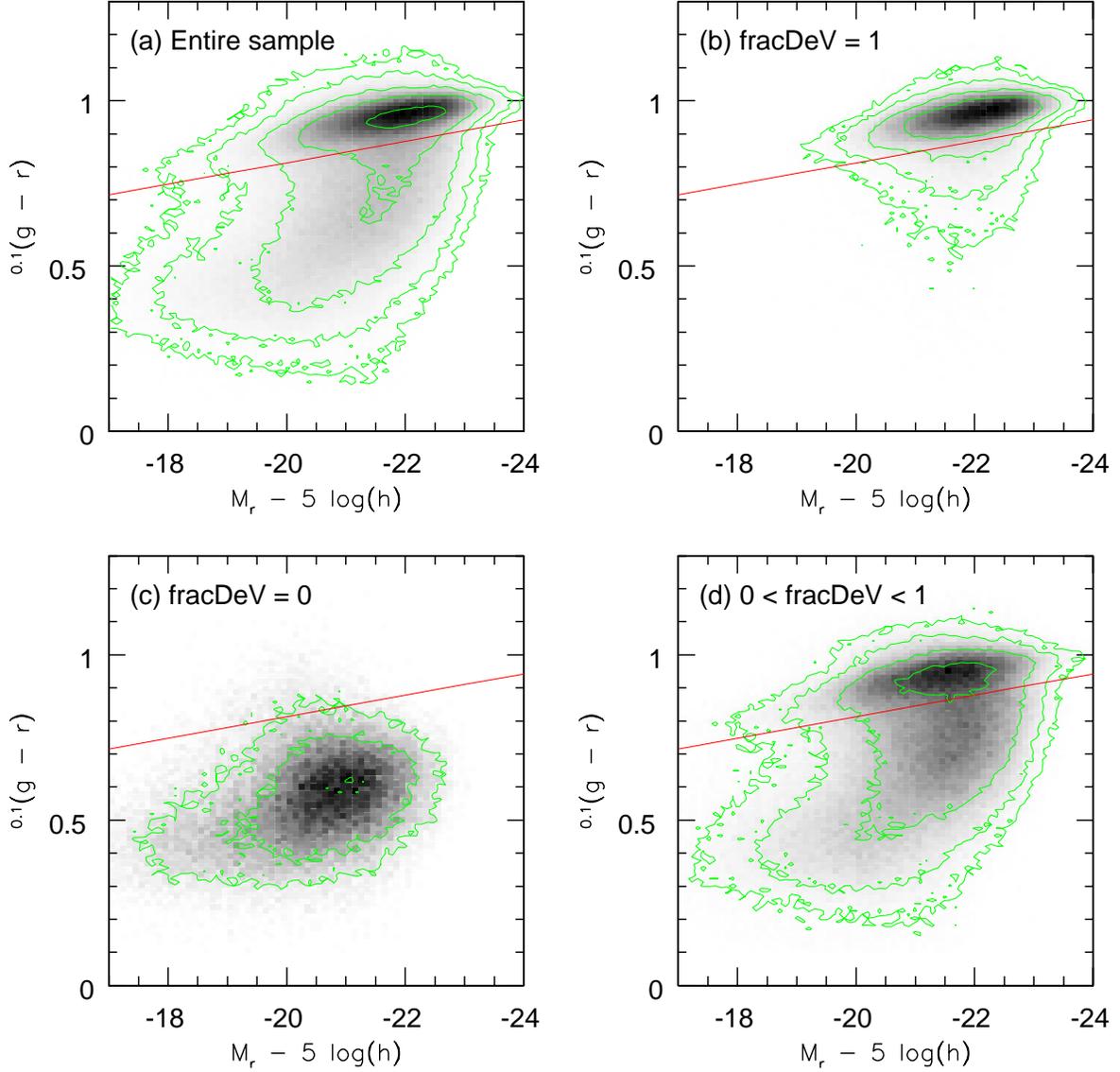} 
\caption{\label{f:galCMD} Galaxy sample color properties. We adopt the locus
$^{0.1}(g-r)\,=\,0.78\,-\,0.0325\,(M_r\,-\,5\,\textrm{log}\,h\,+\,19)$ of
\citet{bailin_2007} (shown in red) as the color division between blue cloud and
red sequence galaxies. Panel (a): all galaxies in our sample (panels b, c, \& d
combined). Panel (b): pure de Vaucouleurs galaxies ($\FD = 1$). Panel (c): pure
exponential galaxies ($\FD = 0$). Panel (d): combined-profile objects ($0 < \FD
< 1$). The gray scale is normalized seperately in each panel. } 
\end{figure}

Finally, we use this K-corrected photometry to assign each galaxy one of two
color types (red or blue) using the following criteria
\citep{bailin_2008}\footnote{This division deviates slightly from that
presented in \citet{bailin_2008}. We use the values from version 1 of their
arXiv preprint \citep{bailin_2007}. The differences between the two divisions
are minor and do not affect our results.}: 
\begin{eqnarray} 
\textrm{red galaxies:}&&\;\;^{0.1}(g-r) > 0.78 - 0.0325 ( M_r -
5\,\textrm{log}\,h + 19 ) \\
\textrm{blue galaxies:}&&\;\;^{0.1}(g-r) < 0.78 - 0.0325 ( M_r -
5\,\textrm{log}\,h + 19 ) 
\end{eqnarray}
This effectively separates the so-called red sequence and blue cloud (Fig.\
\ref{f:galCMD}). These data products allow us to isolate and characterize the
populations which give rise to the alignment trends we observe.


\section{Group Selection and Alignment Geometry} \label{s:selection}

Before analyzing alignment effects, we must first identify physically-related
galaxies. To do so we separate galaxies into groups. We define a fixed-size
cylindrical ``neighbor volume'' of $2400\kms$ in height and $1\,\hinv\,\mpc$ in
radius (Fig.\ \ref{fig:cylinder_angles}a) centered on each galaxy, within
which we search for companions. Each group consists of one host galaxy (HG) and
at least one satellite galaxy (SG). 

Initially, any galaxy in our data set is a potential HG. We begin the selection
process by eliminating all galaxies whose neighbor volume intersects either an
SDSS spectroscopic survey border or the limiting redshift of our sample. For
each remaining object, we produce a list of companions (all galaxies within the
neighbor volume), eliminating any solitary hosts. Next, using the bolometric
luminosity calculated with the LRT code (\S\ref{s:data}), we eliminate all
galaxies which have a more luminous companion within the neighbor volume. The
remaining objects comprise our final HG list.

Our group sample consists of all companions (SGs) within the neighbor volume of
all HG list members. In total, we find 291,435 SGs in the vicinity of 101,850
unique HGs. In our implementation, SGs can belong to multiple groups while the
HGs cannot.

\subsection{Alignment Angles} \label{ss:angles}

\begin{figure}
%
\plottwo{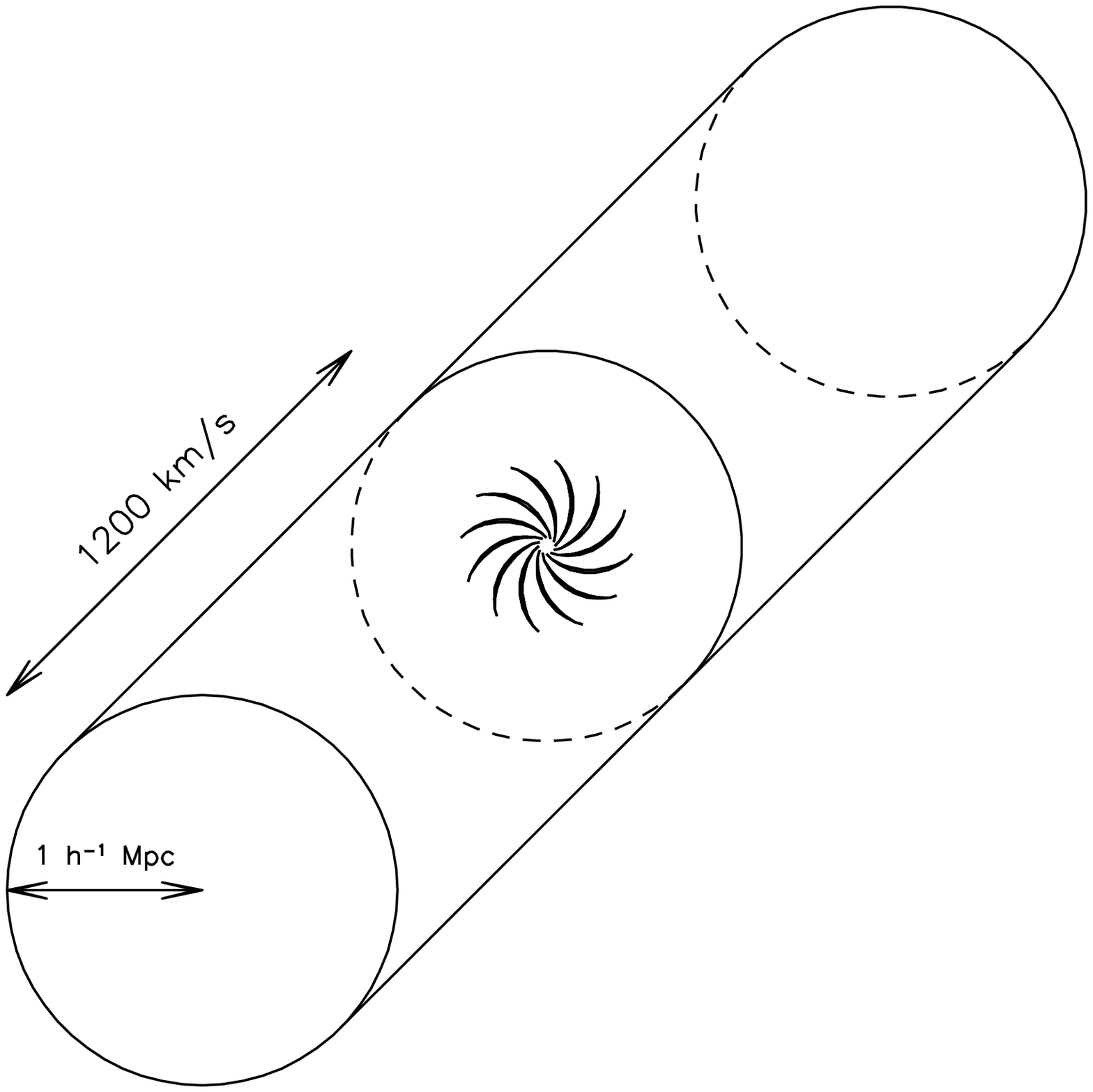}{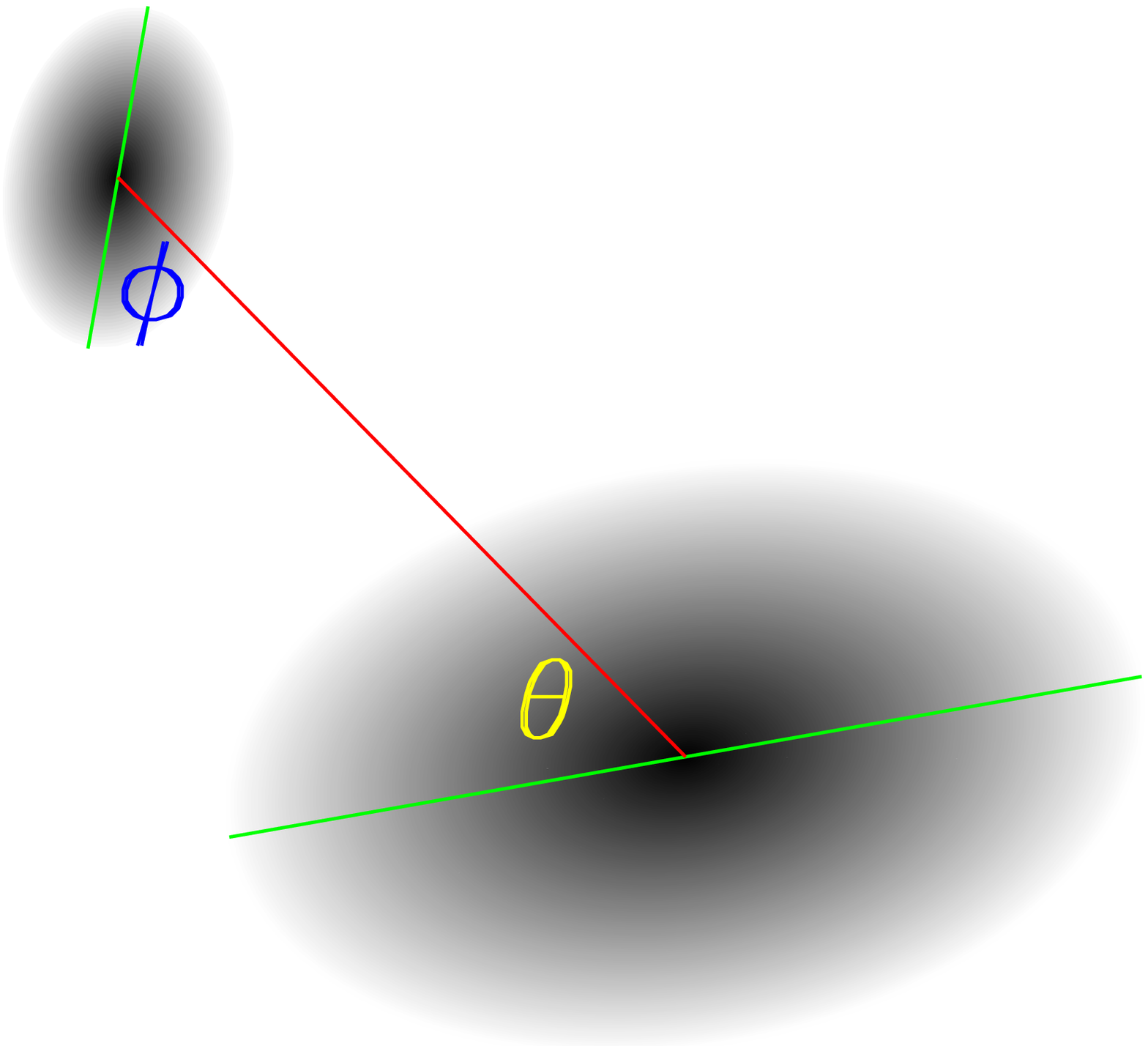}
\caption{\label{fig:cylinder_angles} Left: Neighbor volume cylinder diagram.
Satellite galaxies have $\DR<1\hmpc$ and $\DV<1200\kms$ relative to their host
galaxies. Right: schematic diagram illustrating the angles $\theta$ and $\phi$
we use to characterize the relative positions and orientations of host and
satellite galaxies. The angle $\theta$, also called location angle, measures
the angular position of a satellite galaxy relative to the direction of the
host galaxy major axis. $\phi$ measures the orientation of the satellite major
axis relative to ``hostward'' (radial). }
\end{figure}

Several distinct alignment trends have been observed to date (see
\S\ref{s:intro}). Identifying and quantifying these effects requires more than
one angular separation measurement. For each HG-SG pair, we compute a location
angle ($\theta$) and radial alignment angle ($\phi$). These quantities
characterize the geometries we investigate (see Fig.\
\ref{fig:cylinder_angles}). For simplicity we have adopted the symbols of
\citet{falten1}. We separately calculate each using isophotal, exponential, and
de Vaucouleurs PA values. To exploit inherent symmetries, we reduce all angles
to values between $0\DEG$ and $90\DEG$. 

We define $\theta$ to be the angle between the HG major axis and the line
connecting the centers of host and satellite (i.e., the host-satellite
separation vector). This quantity describes the location of a companion galaxy
with respect to the major ($\theta = 0\DEG$) or minor ($\theta = 90\DEG$) axis
of its host (see Fig.\ \ref{fig:cylinder_angles}). We use this to characterize
anisotropy in the satellite galaxy distribution (i.e., the Holmberg Effect). 

Calculating $\theta$ accurately requires a robust measurement of the HG PA. To
this end, we include only those HGs with $\qiso \le 0.9$ and $\qmom \le 0.9$.
This shape requirement eliminates 78,070 galaxy pairs (27\%). We also exclude
all objects with $\DPA > 15\DEG$. By itself, this cut eliminates 90,341 pairs
(31\%). In combination, the shape and $\DPA$ cuts eliminate 43\% of our group
sample but ensure that the remaining galaxy pairs (166654 SGs around 58259
unique HGs) have usefully high HG PA accuracy and thus reliable $\theta$
measurements. This procedure maximizes the sample size without risking
inclusion of spurious PA measurements.

Our second angular separation, $\phi$, is the projected angle between satellite
galaxy position angle and the projected host-satellite separation vector (Fig.\
\ref{fig:cylinder_angles}). We use $\phi$ to expose any preferred direction in
SG elongation. The orientation extrema are radial ($\phi = 0\DEG$, directly
toward the host) and tangential ($\phi = 90\DEG$). 

To properly measure $\phi$, we require a robust measurement of SG PA. We thus
include only those SGs with $\qiso \le 0.9$ and $\qmom \le 0.9$. Applied to
SGs, this shape cut eliminates 66,159 pairs (23\%). We also eliminate SGs that
lack good agreement between model and isophotal PA (i.e.\ SG $\DPA > 15\DEG$).
On its own, this $\DPA$ requirement eliminates 85,105 galaxy pairs (29\%).
Together, these criteria reduce our sample size by 38\% (to 180,777 SGs around
77,213 unique HGs) but minimize the inclusion of spurious PA measurements. We
do not apply any shape cuts to HGs in measuring $\phi$. 

For each object, we calculate angular separations using spherical trigonometry,
adopting the galaxy centers indicated by the SDSS J2000 coordinates. Position
uncertainty is negligible for our purposes. SDSS determines centroids (the
first moment of light distribution) using an adaptively-smoothed,
PSF-length-scaled quartic interpolation algorithm \citep{SDSS_Astrometry} which
is accurate to within a few tens of milliarcseconds. The greatest sources of
uncertainty in calculating $\theta$ and $\phi$ are due to the PA measurements.
We follow the shape and $\DPA$ prescriptions outlined above to minimize the
potential impact of ambiguous, uncertain, or contaminated PA determinations on
our results.

\section{Statistical Sample Properties} \label{s:sample}

Before the cuts on HG ellipticity and position angles, the group sample
consists of 291,435 satellite galaxies (SGs) within the cylindrical neighbor
volume of 101,850 unique host galaxies (HGs), indicating a mean SG count of
2.86 per system. Nearly half of systems (48,018) are galaxy pairs and 99\% of
systems contain 20 or fewer SGs. The largest observed group contains 169
satellites (Abell 2197). 

\begin{figure} 
%
\plotone{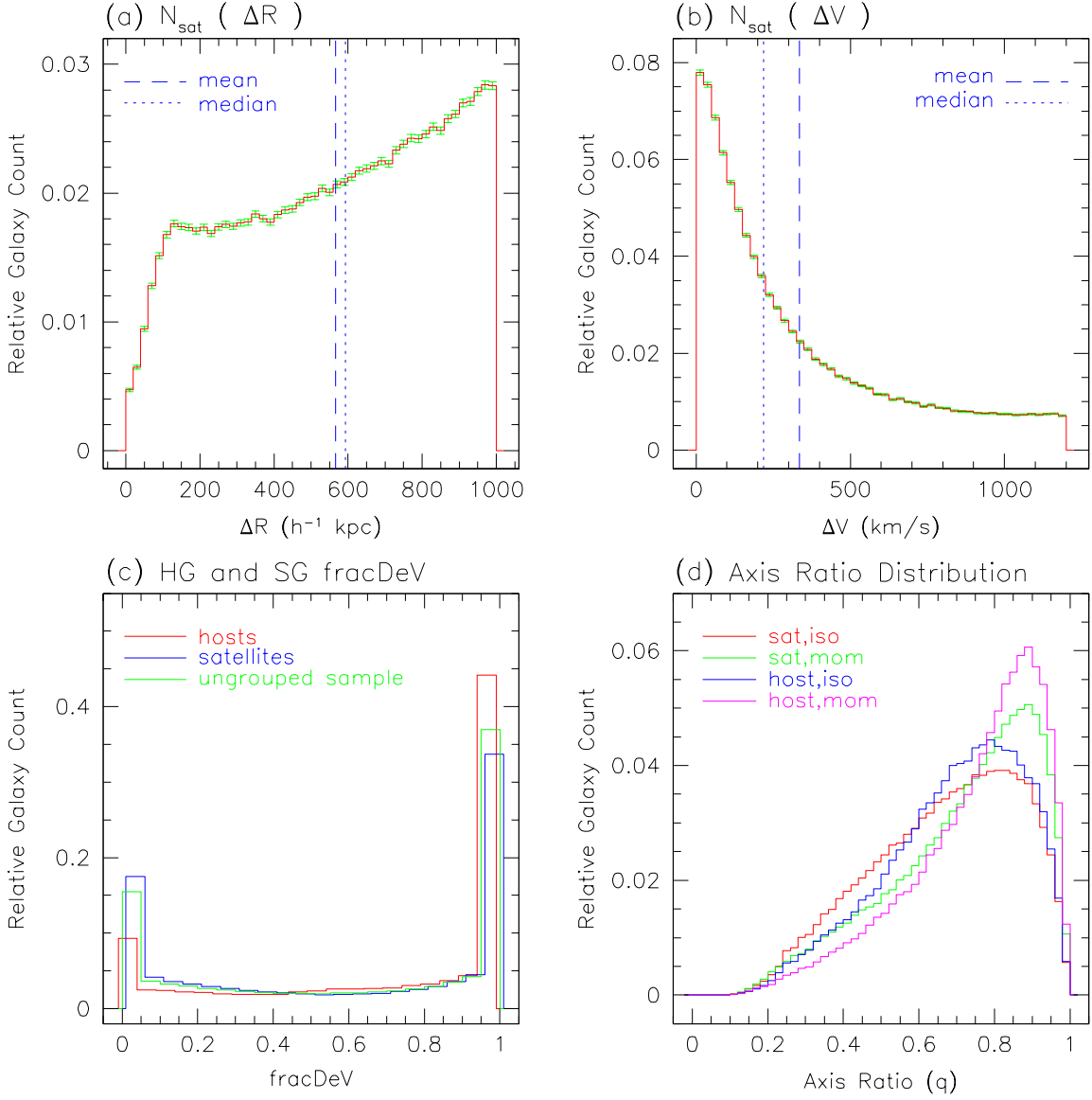}
\caption{\label{f:standard_sample} Properties of the galaxy group sample. Panel
(a): satellite galaxy (SG) counts as a function of projected radius ($\DR$).
Panel (b): SG density as a function of velocity separation ($\DV$). Panel (c):
host galaxy (HG) and SG $\FD$ relative to the ungrouped data set. Panel (d):
isophotal and moments axis ratios ($\qiso$, $\qmom$) for both HG and SG
populations. }
\end{figure}

In Figure \ref{f:standard_sample} we present spatial, velocity, concentration,
and shape distributions of objects in our sample of grouped galaxies. Figure
\ref{f:standard_sample}a shows the number of SGs within annular bins of
projected radial separation ($r^2 dN/dr$, where $r$ is the HG-SG projected
radial separation and $N$ is the number of SGs per bin). Although this metric
does not properly portray the SG spatial density, it serves to illustrate the
rapid fall in galaxy counts at small projected separations ($\DR \lesssim
150\hkpc$) due to the fiber collision effect. In reality, of course, galaxy
density ($dN/dr$) is a steeply dropping function of separation. Satellite
density also falls off steeply with increasing velocity separation ($\DV$,
Fig.\ \ref{f:standard_sample}b).

Figure \ref{f:standard_sample}c compares the $\FD$ distribution of HGs and SGs
in our group sample to that of our entire data set which also includes
ungrouped galaxies, i.e.\ isolated galaxies or galaxies near the survey
boundary. Relative to the ungrouped objects, HGs are more likely to have high
$\FD$ ($>\,0.9$) and slighly less likely to have very low $\FD$ ($<0.1$). SGs
in groups show the opposite trend: there are fewer high $\FD$ ($>0.9$) galaxies
and more low $\FD$ ($<0.1$) galaxies relative to the ungrouped sample. Neither
HGs nor SGs deviate noticeably from the parent population for $0.1<\FD<0.9$.
Lastly, Fig.\ \ref{f:standard_sample}d compares isophotal and moments-based
axis ratios of HGs and SGs. We find (in order of increasing ellipticity) that
$\hqmom>\sqmom>\hqiso>\sqiso$. We obtain this same order using both mean and
median $\qiso$ and $\qmom$. Using $\qmom$ we also find, however, that a larger
fraction of SGs exhibit extreme elongation ($q<0.2$) than HGs.

\section{Results} \label{s:results}

\subsection{Analysis Methods and Significance} \label{ss:significance}

To date, most galaxy alignment studies have focused on trend detection rather
than characterization. We wish to verify and extend these previous analyses by
quantifying the strength of all alignment signals we find. To that end, we
employ the following procedure. Within our grouped galaxy catalog, we apply
various cuts in different observables (e.g. color, $\DV$, $\DR$ and $\FD$). The
resulting subsamples are binned into histograms to which we apply weighted
linear least-squares fits. Fit parameters provide measures of both trend
strength (line slope) and significance (slope uncertainty). We then verify our
estimates of the significance of the trends we find by comparing to simple
analytic estimates. In quantifying the magnitude of alignment effects, we are
better equipped to isolate the galaxy populations responsible for these
alignments.

Our histogram-based analysis affords several advantages. Firstly, linear fits
to binned data are simple to perform and provide very reliable results with our
typical subsample sizes. Because of this simple approach, an analytic estimate
of the statistical significance of any alignment trend is straightforward to
calculate. More importantly, by capturing the shape of the histogram in a
single value (the fractional slope), we are able to concisely investigate
alignment signal strength in the multidimensional parameter space of galaxy
attributes. Fortunately, all of the alignment trends we observe are very well
described by a linear trend. Lastly, the histograms we produce are visually
similar to, and thus directly comparable to, the results of many previous
studies.

For this analysis, we define a galaxy alignment trend as a systematic variation
in the linear density of galaxies (the histogram) as a function of either
$\theta$, the angular position of SGs relative to the HG major axis, or $\phi$,
the orientation of the SG relative to the HG-SG separation vector
(\S\ref{ss:angles}). A flat (zero slope) histogram implies no alignment signal
(isotropy) whereas a very steep slope constitutes a strong alignment effect. We
use fractional measurements to better compare alignment trends in galaxy
subsamples of different sizes.

Our trend-fitting procedure operates as follows. First, we define a subsample
of grouped objects with the desired criteria. We then produce a histogram by
binning these objects in separation angle. Assuming Poisson errors, each bin
$i$, which contains $N_i$ galaxies, has an uncertainty $\sigma_i = \sqrt{N_i}$.
Dividing each bin and uncertainty by the bin width yields galaxy line density
($\rho_i$) in each bin. We obtain the best-fit slope ($A$), its uncertainty
($\sigma_A$), and intercept ($B$) with a weighted least-squares ($\chi^2$
minimization) fit to these density values. From these values we compute the
fractional slope ($A / B$) and alignment significance ($A / \sigma_A$). For
convenience, we convert these to percent galaxy density change across the
angular interval [$0,90\DEG$],
\begin{equation} 
\Delta N (\%) = 100\% \; (90\DEG A / B)
\end{equation} 
and its uncertainty,
\begin{equation} 
\sigma_{\Delta N} (\%) = 100\% \; (90\DEG \sigma_A / B)
\end{equation} 
By computing the above values in different galaxy subsets, we can quickly
identify where in parameter space a given effect originates and whether or not
variations between subsamples are statistically significant. 

To find the source of an alignment effect, we observe how alignment strength
and significance vary when the parent population is split into subsamples. For
each galaxy attribute we investigate (e.g., $\FD$, color), we divide the group
sample into different bins according to that attribute. If a particular
subpopulation is responsible for the observed alignment trend, its alignment
strength will be diluted by the inclusion of galaxies from other populations.
If, in dividing the group sample, we improve our selection accuracy (i.e., we
create a subsample wherein a larger fraction of galaxies produce the alignment
trend), we observe an increase in alignment strength (steepening of histogram
slope). If, however, the resulting subsamples all have similar strength to each
other or to the parent population, then the alignment effect does not depend on
that particular attribute used to subdivide them. By repeating this procedure
we can identify what galaxy subpopulations do and do not exhibit different
types of alignment.

\begin{figure} 
%
\plotone{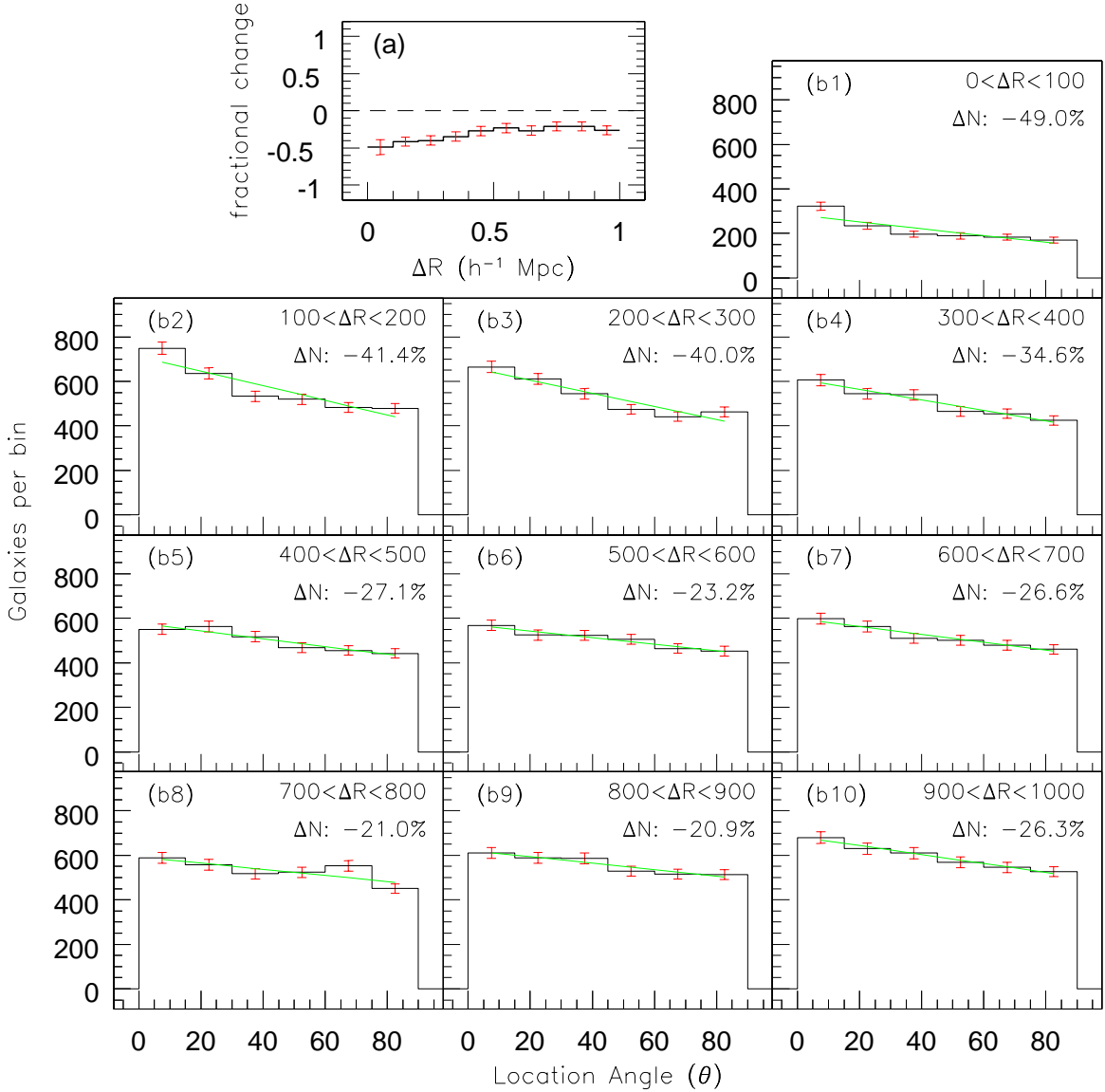}
\caption{\label{f:mondo_ex} Illustration of our histogram condensation analysis
technique. The original subsample, consisting of hosts and satellites with high
concentration ($\FD>0.9$), is further subdivided into 10 bins of radial
host-satellite separation ($100\hkpc$ per bin, panels b1-b10). We fit a linear
model to this histogram within each subdivision and compute the fractional
change and its uncertainty. These fractional changes are then reproduced in the
panel (a) as a function of radial separation. The error bars are the slope
uncertainties from the fitting procedure. }
\end{figure}

In representing individual histograms by a single number (the fractional galaxy
density change), we are able to expand our histogram-based analysis to multiple
parameters simultaneously. We illustrate this technique in Figure
\ref{f:mondo_ex}. The final result (Fig.\ \ref{f:mondo_ex}a) divides the
subsample of highly concentrated host and satellite galaxies ($\FD>0.9$) into
10 bins of radial separation ($100\hkpc$ per bin). We separately fit the galaxy
density histogram and compute the fractional change within each subdivision
(Figures \ref{f:mondo_ex}b1 - \ref{f:mondo_ex}b10) and plot these findings as a
funtion of radial separation (Fig.\ \ref{f:mondo_ex}a). The error bars are in
the condensed result panel are the slope uncertainties from the individual
histogram fits. Applying this technique for many different subsamples of
observable host and satellite galaxy properties (e.g., concentration or color)
allows us to identify the more complex parameter dependence of certain
alignment trends.

For a given subsample with $N_{tot}$ galaxies, the following analytic estimate
determines the expected alignment significance:
\begin{equation}
\frac{S}{N} \equiv \left ( \frac{A}{\sigma_{A}} \right ) = \sqrt{\frac{4}{3}}
\, \frac{A \, (\Delta \theta)^2}{\sqrt{N_{tot}}},
\end{equation}
where $\Delta \theta = 45\DEG$ is the half width of the possible range of
alignment angles (see Appendix \ref{a:SN_derive}). We find excellent agreement
between this estimate and the significance determined explicitly by our fitting
procedure. 

We now summarize our results for the major axis preference and hostward
elongation. We compare our results with those from the literature in
\S\ref{s:comparison}.


\subsection{Major Axis Preference} \label{ss:results_map}

The most noticeable anisotropy we find is a satellite galaxy (SG) distribution
which favors the host galaxy (HG) projected major axes in projection. For the
full group sample we observe a $-12.4\pm0.8\%$ change in galaxy density between
major and minor axis at $15\sigma$ significance (black line in Figures
\ref{f:d1_hg_fdev}-\ref{f:d1_scount}). As we will now show, this anisotropy is
strongest for red, centrally-concentrated SGs of red, centrally-concentrated
HGs. Further, the effect appears to increase in strength and significance with
decreasing physical (3-D) separation.

Selecting host-satellite pairs by host $\FD$ sheds more light on the origin of
this effect (Fig.\ \ref{f:d1_hg_fdev}). While our catalog as a whole exhibits
this anisotropy at the $15\sigma$ level, we find that it arises entirely from
systems with HG $\FD>0.5$. More specifically, the signal is strongest among
systems with HG $\FD>0.9$ which show a $-23.9\pm1.1\%$ SG density change from
major to minor axis at $21.2\sigma$. Galaxy groups with $0.5<\HG\,\FD<0.9$ show
a weaker preference ($-6.8\pm1.6\%$, $4.3\sigma$). At the same time, our fit to
this group shows a 3.6-fold decrease in significance relative to the total
galaxy population. We see no significant trend for HG $\FD<0.5$.

Splitting the group sample by SG $\FD$, we observe a decreased range of
alignment strengths (Fig.\ \ref{f:d1_sg_fdev}). The two de Vaucouleurs
profile-dominated SG subsamples (SG $\FD>0.9$ and $0.9>\SG\,\FD>0.5$) show
enhanced alignment strength ($-17.2\pm1.3\%$ and $-15.9\pm1.8\%$) relative to
the total population and are consistent with each other. The two exponential
profile-dominated SG subsamples ($0.1<\SG\,\FD<0.5$ and SG $\FD<0.1$) are also
consistent with each other (changes of $-7.1\pm1.8\%$ and $-6.2\pm1.7\%$,
respectively) but shallower than the combined group sample.

Breaking the sample down by HG and SG color shows a similar picture. We observe
a very strong effect between red HG - red SG pairs ($-26.2\pm1.3\%$,
$19.5\sigma$, Fig.\ \ref{f:d1_color}). Blue SGs around red HGs show a less
significant alignment trend ($-14.3\pm1.5\%$, $9.6\sigma$). We find no evidence
of anisotropy among either red or blue SGs of blue HGs. These latter two
subsamples are consistent with each other ($-1.5\pm1.8\%$ and $-2.6\pm2.2\%$
decreases) and with isotropy. We deduce that red and centrally-concentrated
(i.e.\ elliptical) HGs and SGs are almost exclusively responsible for the major
axis alignment preference.

We observe this major axis preference across a wide range of HG-SG velocity
separation (Fig.\ \ref{f:d1_deltaV}). Although it is most significant for
$|\DV|<300\kms$ ($-14.8\pm1.0\%$, $14.3\sigma$), we observe an equally strong
($14.8\%$ decrease) trend for $300\kms<|\DV|<600\kms$ albeit at somewhat
reduced significance ($8.3\sigma$) due to the smaller subsample size. The
subsamples at greater velocity separations ($\DV>600\kms$) are consistent with
each other and with isotropy. We find a similar effect in projected radial
separation (Fig.\ \ref{f:d1_radius}). Alignment is strongest ($-20.4\pm2.0\%$,
$10\sigma$) for the closest SGs ($\DR<250\hkpc$) but is still detected at
nearly $6\sigma$ to the group size limit of $1\hmpc$. Taken together, these
velocity and radial separation trends indicate an increasing spatial anisotropy
in favor of projected HG major axis with decreasing 3-D separation.

We also investigate anisotropy as a function of (spectroscopic) galaxy group
size (Fig.\ \ref{f:d1_scount}). We find that alignment strength increases
steadily with increasing group size. Among SGs from the smallest groups
($1<N_{SG}<5$; 95186 objects) we observe a $-9.6\pm1.1\%$ galaxy density change
towards the minor axis at $9\sigma$ significance. In the largest groups
($N_{SG}>20$; 16122 objects) we find a $-23.5\pm2.4\%$ change ($9.6\sigma$).
Interestingly, the latter subsample exhibits the strongest alignment and
highest significance despite having the fewest galaxies. From this we deduce
that major axis preference exists for all group sizes but is strongest for the
largest galaxy groups. 

\begin{figure} 
\epsscale{0.95}
\plotone{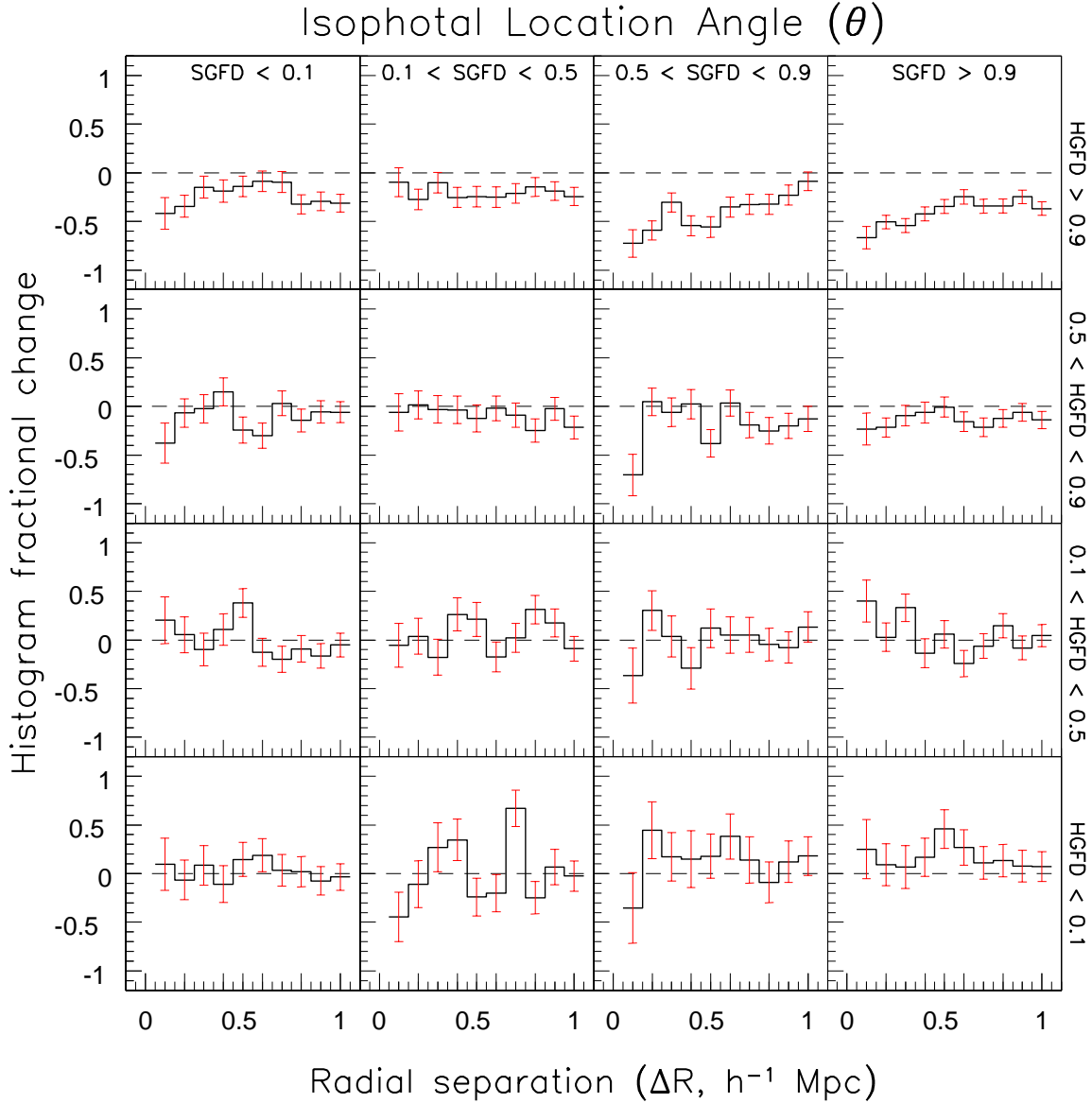}
\epsscale{1.00}
\caption{\label{f:real1i} Differential distribution of satellite galaxy (SG)
angular position ($\theta$) relative to host galaxies (HGs). Each panel column
contains a different range of satellite galaxy fracDeV. Similarly, each panel
row hosts a different range of host galaxy $\FD$. Projected radial separation
($\DR \le 1\hmpc$), in 10 bins of width $100\hkpc$, spans the horizontal axis.
See Fig.\ \ref{f:mondo_ex} for an example of this plotting technique. High
significance (fractionally small error bars) detections of SG preference for HG
major axes (values below 0) are markedly concentrated towards the upper-right
(high $\FD$). All objects have velocity separation $\DV < 500\kms$ and good
agreement between isophotal and model position angles ($\DPA \le 15\DEG$). } 
\end{figure}

In Figure \ref{f:real1i} we explore the interdependence of physical separation
and light profile type on the observed alignment trend. We limit velocity
separation to $|\DV|<500\kms$ and explore the dependence of alignment strength
on projected radial separation ($\DR$). We find that when both HG and SG have
high $\FD$ (especially $\FD>0.9$, top-right panel), the anisotropy strength
(fractional slope) increases with decreasing $\DR$. We conclude that the
dependence of anisotropy on physical separation occurs largely among the same
red, centrally-concentrated galaxy pairs described above. We further discuss
the significance of these results in \S\ref{ss:comp_map}.


\subsection{Hostward Elongation} \label{ss:res_elong}

We find that satellites tend to elongate towards their host galaxies. For the
full group sample, we observe a $-5.7\pm0.8\%$ change (at $7\sigma$) between
the frequency of satellites that exhibit perfect radial (hostward) alignment
relative to those that exhibit tangential alignment (black histogram in Figures
\ref{f:d3_sg_fdev}-\ref{f:d3_scount}). Of the galaxy properties we examined,
the degree of alignment is most strongly dependent on SG $\FD$ (concentration,
Fig.\ \ref{f:d3_sg_fdev}). SGs with $\FD>0.9$ (52,817 objects, 29\% of sample)
exhibit a $-13.9\pm1.4\%$ density change favoring direct hostward alignment at
$9.9\sigma$. None of the other SG $\FD$ subsamples exhibit significant
alignment.

We present hostward elongation ($\phi$) for different HG profile types in
Figure \ref{f:d3_fdev}. Satellites of high-$\FD$ HGs (HG $\FD>0.9$, 94,716
objects, 52\% of sample) show a $-5.8\pm1.1\%$ change in density at
$5.3\sigma$. Satellites of HGs with $0.5<\FD<0.9$ show only a marginal
($3.9\sigma$) $-6.2\pm1.6\%$ change. The slopes of the trends seen in all four
bins are statistically consistent with the total group sample, suggesting that
hostward elongation is largely independent of HG $\FD$.

Galaxy colors (Fig.\ \ref{f:d3_color}) have some effect on alignment strength.
Red SGs around red HGs (56,421 objects, 31\% of sample) show a $-9.8\pm1.4\%$
density change at $7\sigma$. Red SGs around blue HGs exhibit a slightly steeper
trend of $-11.3\pm1.9\%$ at somewhat lower significance ($5.8\sigma$). These
two subsamples are statistically consistent, providing further evidence that
hostward elongation is principally dependent on SG properties. We do not
observe a statistically significant alignment trend among blue SGs of either
blue or red HGs.

There is no evidence that the degree of hostward elongation depends on velocity
separation (see Fig.\ \ref{f:d3_deltaV}). All four subsamples are essentially
statistically consistent with the full group sample. In projected radial
separation (Fig.\ \ref{f:d3_radius}), we find marginal ($5\sigma$) evidence for
hostward elongation in both the largest $\DR$ and smallest $\DR$ bins. At large
separation ($750\hkpc<\DR<1\hmpc$) we find a $-7.2\pm1.4\%$ density change at
$5.3\sigma$. In the innermost radius bin ($\DR<250\hkpc$), we find a
$-9.7\pm1.9\%$ density change at $5.1\sigma$. These results suggest that 3-D
physical separation plays a mininal role in the degree of hostward elongation.

Lastly, we examine the influence of group size on hostward elongation (Fig.\
\ref{f:d3_scount}). All four group size subsamples are consistent with the
whole group sample, suggesting that hostward elongation is independent of group
size.

\section{Caveats and Complications} \label{s:caveats}

\subsection{Discrepant Position Angles} \label{ss:discrepant_PA}

Our greatest cause for concern is the accuracy of measured galaxy position
angles (PAs). During our analysis, we identified systematic PA discrepancies
between model- and isophote- based position angles ($\PAD$ and $\PAI$) that
critically affect our results. Some of these differences, primarily those due
to isophotal twisting, are physically genuine. However, in many cases,
isophotal contamination by artifacts and nearby objects appears to be the
culprit. We suspect that these variations may be largely responsible for the
general disagreement between many of the results on hostward elongation
presented in the literature. We detail our findings in the remainder of this
section.

\begin{figure} 
%
\plotone{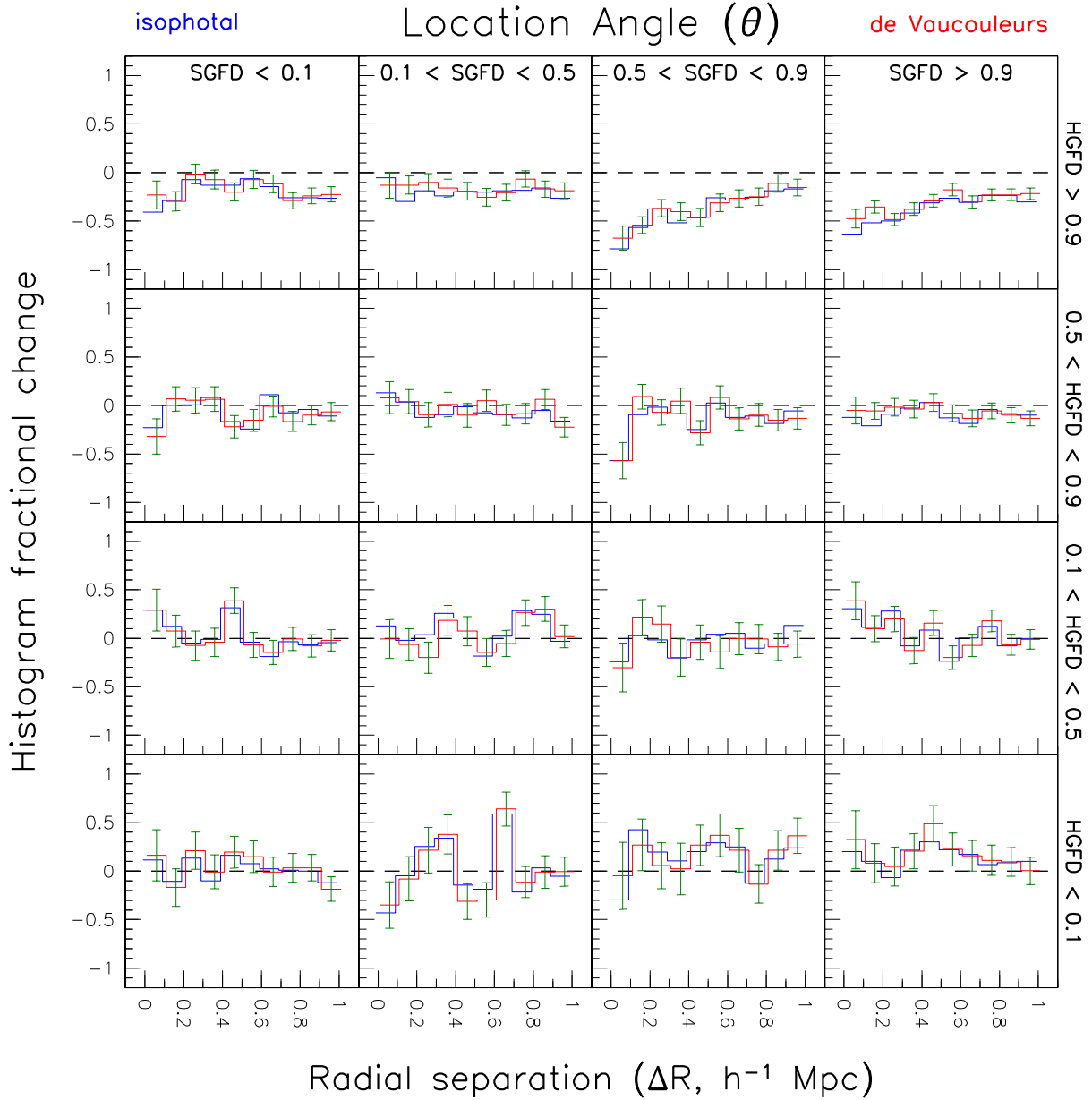}
\caption{ \label{f:iso_dev_LA_comp} Satellite galaxy (SG) distribution
($\theta$) trends of the \textit{same} hosts and satellites using isophotal
position angle (blue) and galaxy model PA (red). We exclude round objects
($\qiso>0.9$ or $\qmom>0.9$) for which PA may be inaccurate.  The isophotal
slope error bars, comparable in magnitude to those of the de Vaucouleurs
slopes, have been omitted for clarity. Comparing these two cases, we observe
little difference for any profile type or radius, suggesting that the HG
population is unaffected by discrepant PAs. }

\end{figure}

To investigate this PA discrepancy, we compare alignment signals of identical
galaxies using $\PAI$ and $\PAD$. In order to illustrate the discrepancy, we do
not enforce agreement between the two measured PA values. We do, however,
continue to require that $\DR>4\prad$. We first examine location angle
($\theta$) which requires accurate host galaxy (HG) PA. Comparing our $\PAI$-
and $\PAD$-based results (Fig.\ \ref{f:iso_dev_LA_comp}), we observe few
significant differences.

\begin{figure}
\epsscale{0.95}
\plotone{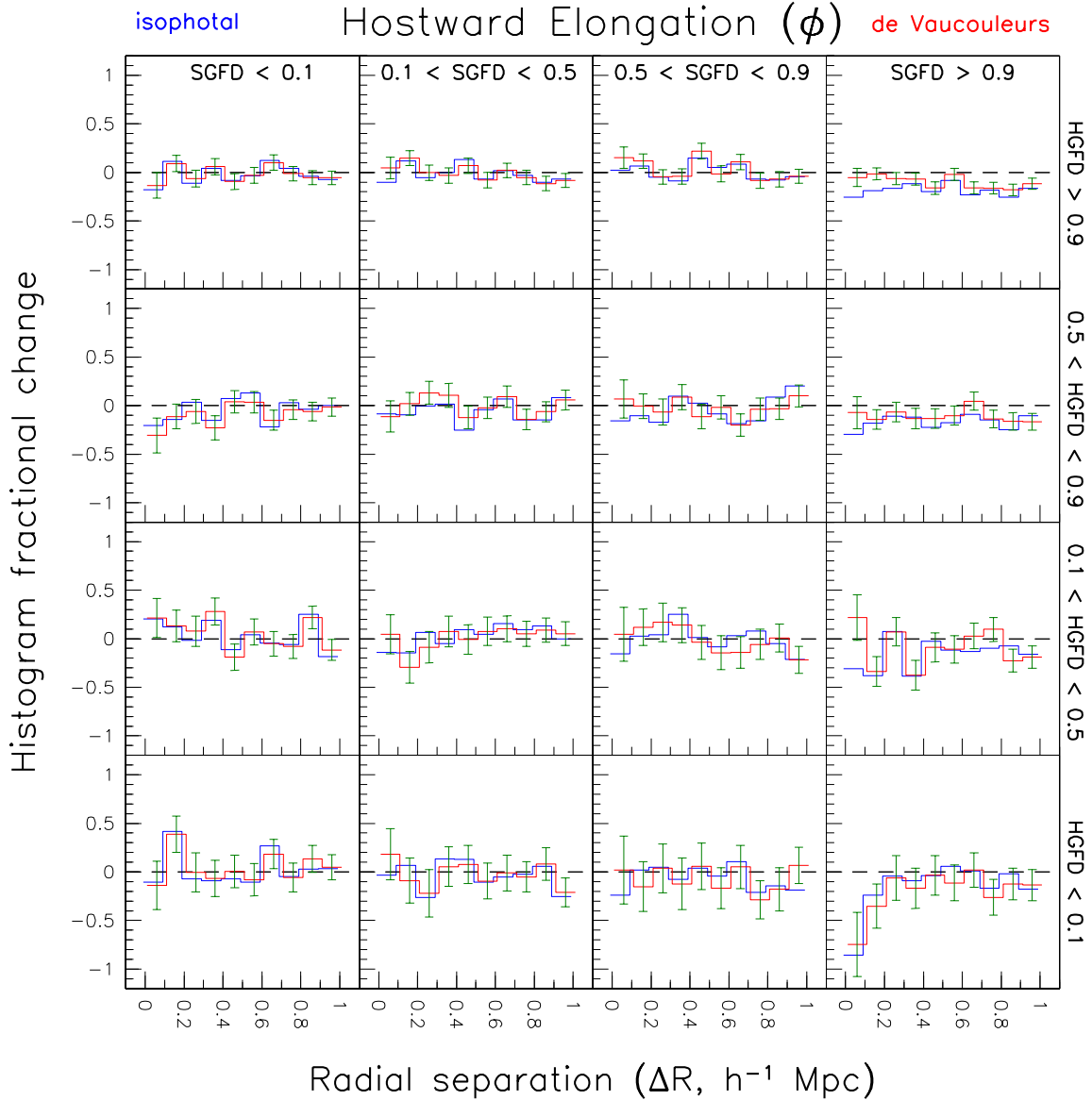}
\epsscale{1.00}
\caption{\label{f:iso_dev_HA_comp} Hostward elongation ($\phi$) in different
bins of host galaxy (HG) $\FD$, satellite galaxy (SG) $\FD$, and projected
radius ($\DR$). The isophotal slope error bars, comparable in magnitude to
those of the de Vaucouleurs slopes, have been omitted for clarity.  We observe
a marked difference between $\phi_{iso}$ (blue) and $\phi_{deV}$ (red) among
the \textit{same} high \FD SGs (rightmost column). In particular, we observe a
striking disparity between isophotal and de Vaucouleurs hostward elongation
among highly concentrated ($\FD>0.9$) SGs of highly concentrated ($\FD>0.9$)
HGs (top-right panel). We examine this subsample in greater detail in Fig.
\ref{f:PA_newcomp}. The discovery of this discrepancy motivated our $\DPA <
15\DEG$ requirement. }
\end{figure}

\begin{figure}
\epsscale{1.00}
\plottwo{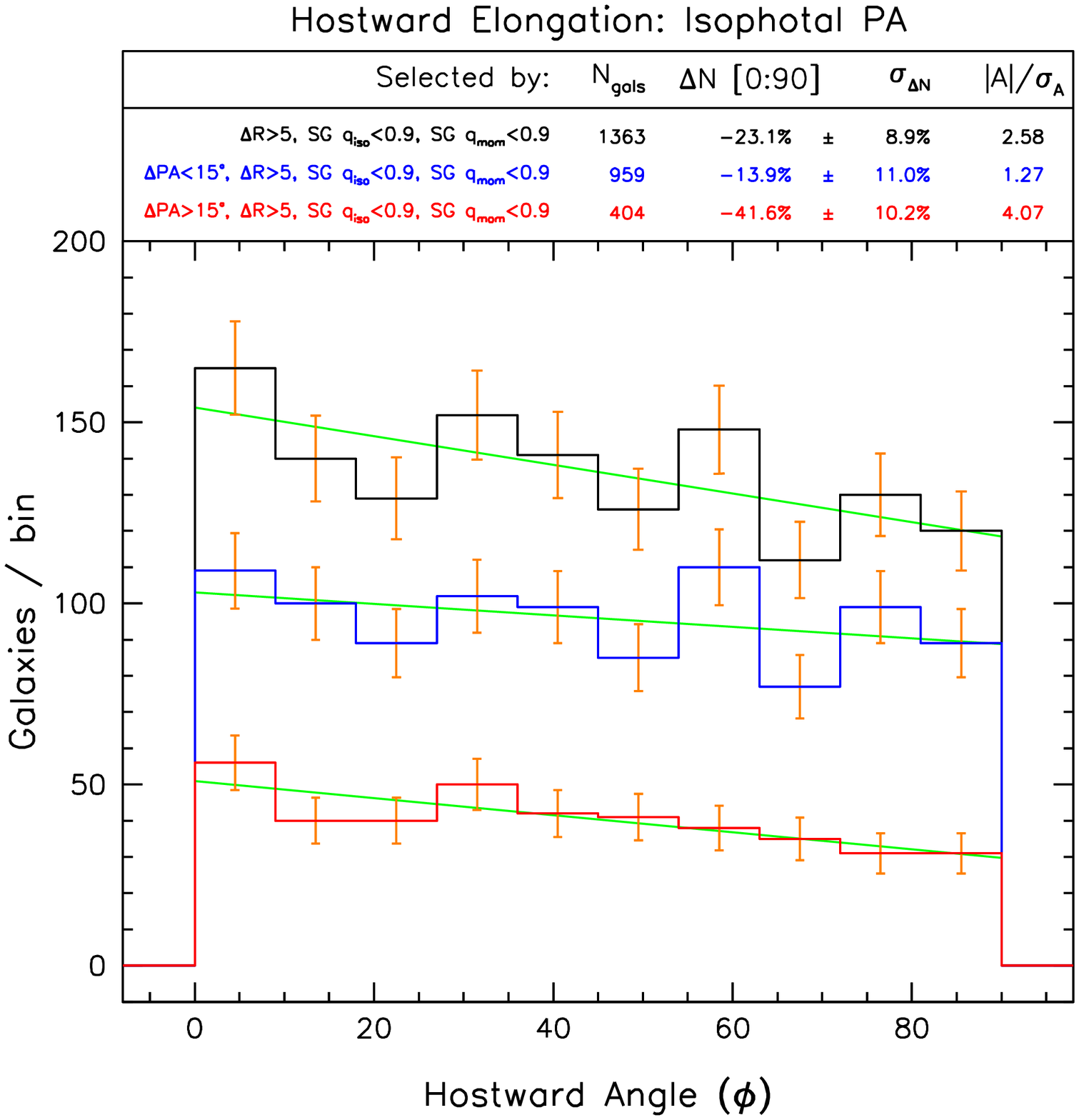}{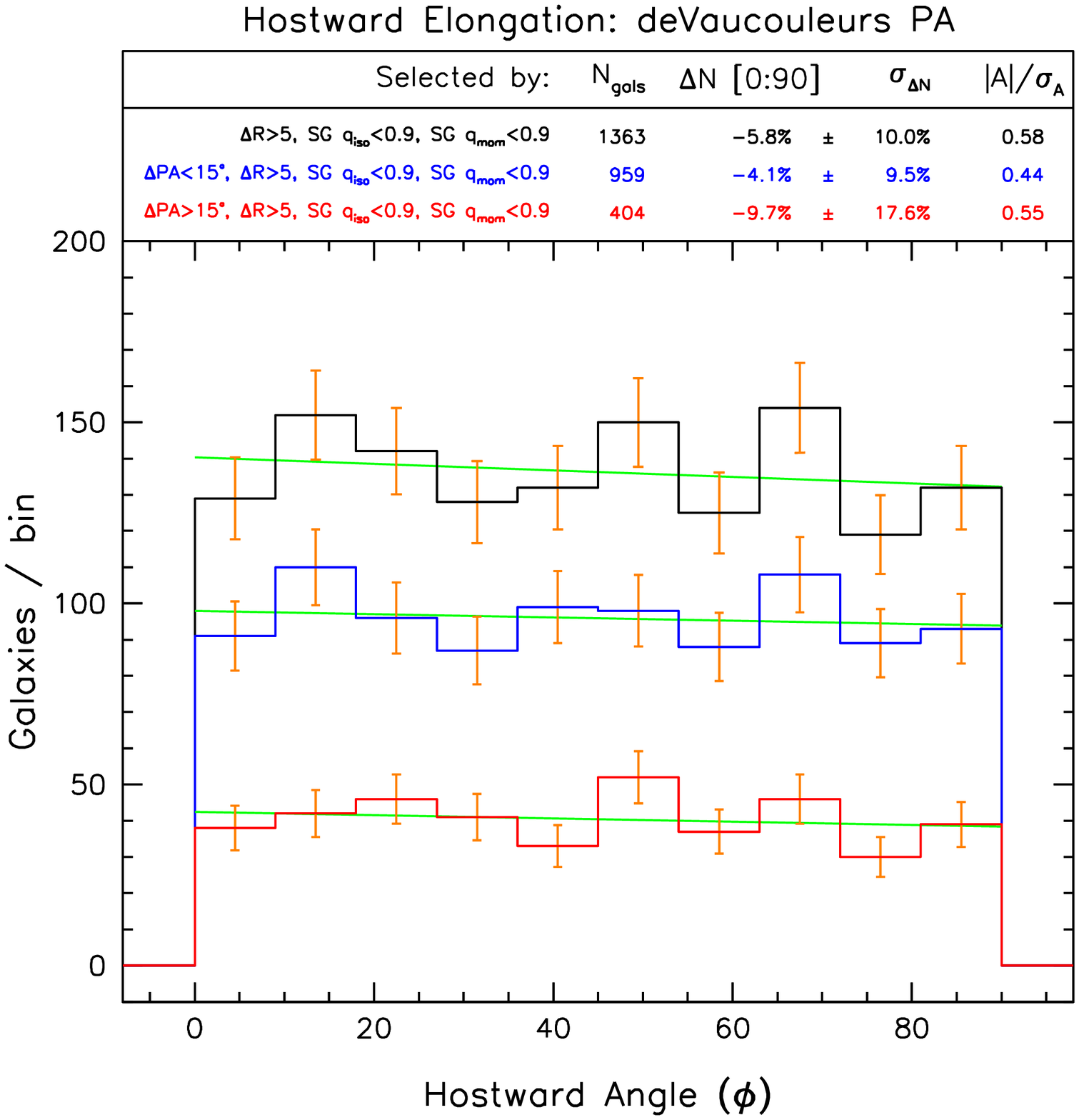}
\epsscale{1.0}
\caption{\label{f:PA_newcomp} Closer inspection of hostward elongation ($\phi$)
in the subsample of host-satellite pairs with HG $\FD>0.9$, SG $\FD>0.9$,
$|\DV|<500\kms$, and $\DR<100\hkpc$, in which we find the greatest discrepancy
between results based on isophote- and de Vaucouleurs model-based PAs (see
Fig.\ \ref{f:iso_dev_HA_comp}). We further separate the galaxies with good PA
agreement ($\DPA<15\DEG$, blue) from those with poor agreement ($\DPA>15\DEG$,
red). The whole subsample (no $\DPA$ requirements, black) is the sum of the red
and blue histograms. In the figures above, we have excluded all satellites
falling within 1 Petrosian radius ($\prad$) of the host galaxy which are very
likely misidentified bright knots within the host itself. Without the
$\DPA<15\DEG$ requirement, the hostward elongation signal from isophotal PA
fractional slope is 4 times that we obtain from de Vaucouleurs model PAs. }
\end{figure}

The situation changes dramatically for the hostward elongation case (Fig.\
\ref{f:iso_dev_HA_comp}). Isophotal and model PAs produce markedly different
results in high $\FD$ SGs (rightmost column), particularly those of high $\FD$
HGs ($\FD > 0.9$, top-right panel). The strong (and high-significance) hostward
alignment we see with $\PAI$ at small projected radii ($\DR < 100\hkpc$) is
absent from $\PAD$-based measurements of \textit{the same satellite galaxies}.
Figure \ref{f:PA_newcomp} compares side-by-side the individual histograms of
this highly discrepant subsample before and after applying the $\DPA<15\DEG$
criterion. After rejecting SGs with $\DPA>15\DEG$ (30\% of objects), both the
alignment trend and discrepancy are significantly reduced. These results
indicate that the apparent degree of hostward elongation is quite sensitive to
PA selection. How best to deal with this depends on the cause of the
discrepancy, which we now investigate.

\begin{figure} 
\epsscale{1.00}
\plotone{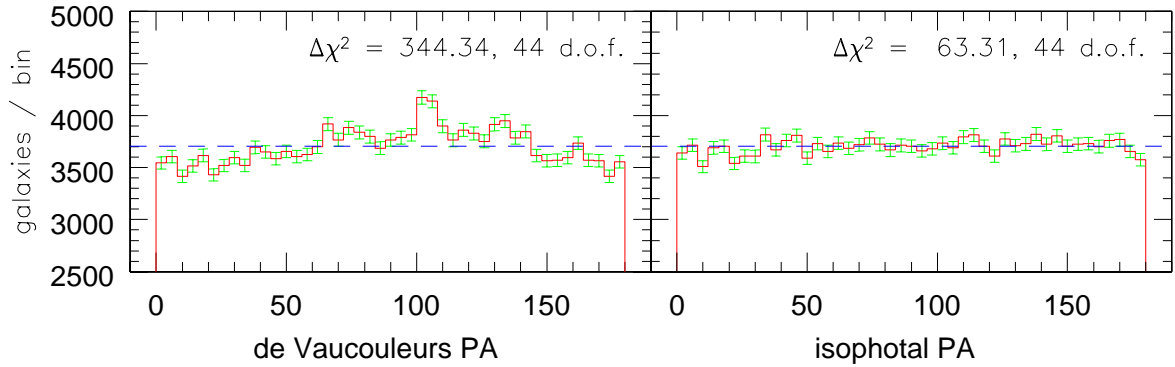} 
\epsscale{1.00}
\caption{\label{f:expPA} The exponential and de Vaucouleurs model-fitting
procedures have several known flaws. Above, the distribution of position angle
(PA) is shown for exponential model fits (left) and isophotes (right). While
the isophotal PAs have an effectively random distribution, the model-based PA
suffers from significant anisotropy. For reference, we provide the $\Delta
\chi ^2$ between the observed distribution and an isotropic (constant) model. }
\end{figure}

Neither PA measure is perfect. The galaxy models (exponential and de
Vaucouleurs) are fitted to each galaxy across a range of radius. These fits
account for seeing effects (\S\ref{s:data}) and may interpolate across
artifacts and substructures as necessary, which reduces the potential for
errors due to contamination. There are, however, several known yet unsolved
problems within the model-fitting routine. Most importantly, the galaxy model
PA distribution is significantly enhanced between $60\DEG$ and $145\DEG$
(nearly East-West in orientation) rather than uniformly distributed in angle
(Fig.\ \ref{f:expPA}). In addition, the derived galaxy scale radii (and thus
the derived axis ratios) exhibit some discretization \citep{DR6}.  Isophotal
PAs, by contrast, are determined by a best-fit ellipse to 25 mag / sq. arcsec
surface brightness around each galaxy. Unlike the matched galaxy models, the
isophotal ellipse fit is sensitive to a narrow and faint surface brightness
region of each galaxy. As such, the isophotal ellipses (and thus their measured
PAs) are more susceptible to contamination by nearby objects and image
artifacts.

To explore the cause of these discrepant PA measurements in individual systems,
we obtained SDSS images of all high-$\FD$ ($\FD>0.9$) SGs hosted by high-$\FD$
HGs with discrepant hostward elongation signals. We visually compared the
SDSS-reported isophotal ellipses and de Vaucouleurs PAs and scale radii with
the galaxy images. We found that the great majority of discrepant PAs could be
attributed to two causes: intrinsic variations in the PAs of isophotes with
radius ($\sim 70\%$) and PA misestimation due to contamination ($\sim 30\%$).
Contamination was caused (in roughly equal numbers) either by a nearby bright
object or by a bright knot somewhere within the galaxy. 

\begin{figure}
%
\plotone{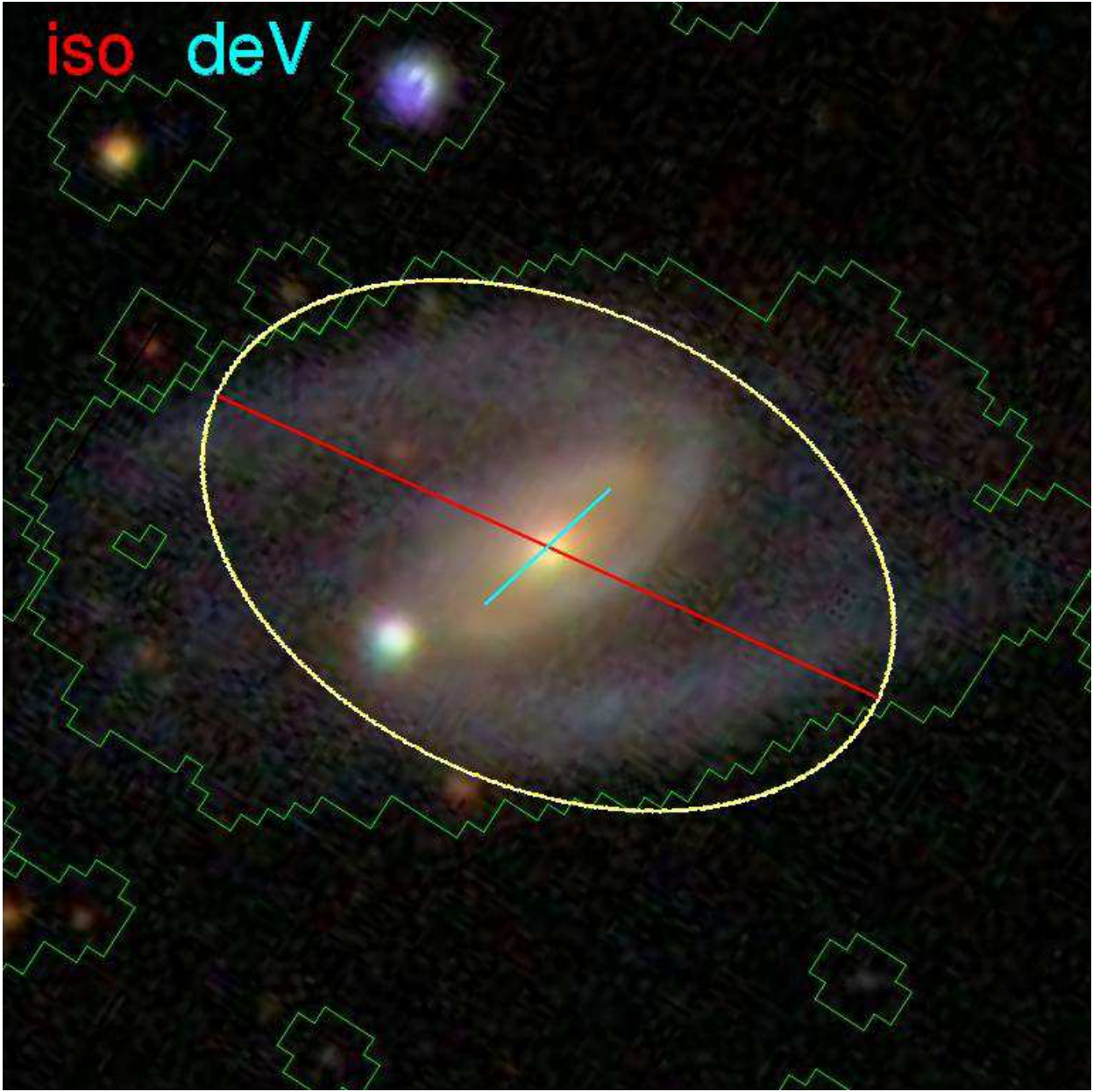}
\caption{\label{fig:PA_RealDiff} Spiral structure causes isophotal twisting,
which can lead to highly discrepant de Vaucouleurs and isophotal position
angles. Above, the galaxy models seek out the light in the bulge and/or bar,
whereas the 25 mag / sq. arcsec isophote roughly detects where the spiral arms
fade into the sky (on much larger scales). } 

\end{figure}

Figure \ref{fig:PA_RealDiff} shows an example of an isolated galaxy for which
the model and isophotal PA differ by nearly $90\DEG$ due to intrinsic isophotal
twisting. Generally, the de Vaucouleurs and exponential galaxy profiles fit to
the brighter galaxy interior while the 25 magnitudes / arcsec$^2$ isophote
traces the outer edge of each galaxy. This is particularly noticeable in this
example, where the de Vaucouleurs model fits primarily the central bar or bulge
region, while the isophotal ellipse traces the outer edge of the spiral arms.
In this fashion, isophotal twisting can lead to highly discrepant PA
measurements. Given that the physical mechanisms which give rise to galaxy
alignments are not well understood, it is not clear which PA determination is
most relevant. This motivated our decision to exclude all galaxies with highly
discrepant ($\DPA>15\DEG$) isophotal and de Vaucouleurs PAs from this analysis.
Further study is required before these interesting objects can be properly
incorporated into an alignment analysis.

\begin{figure} 
%
\plotone{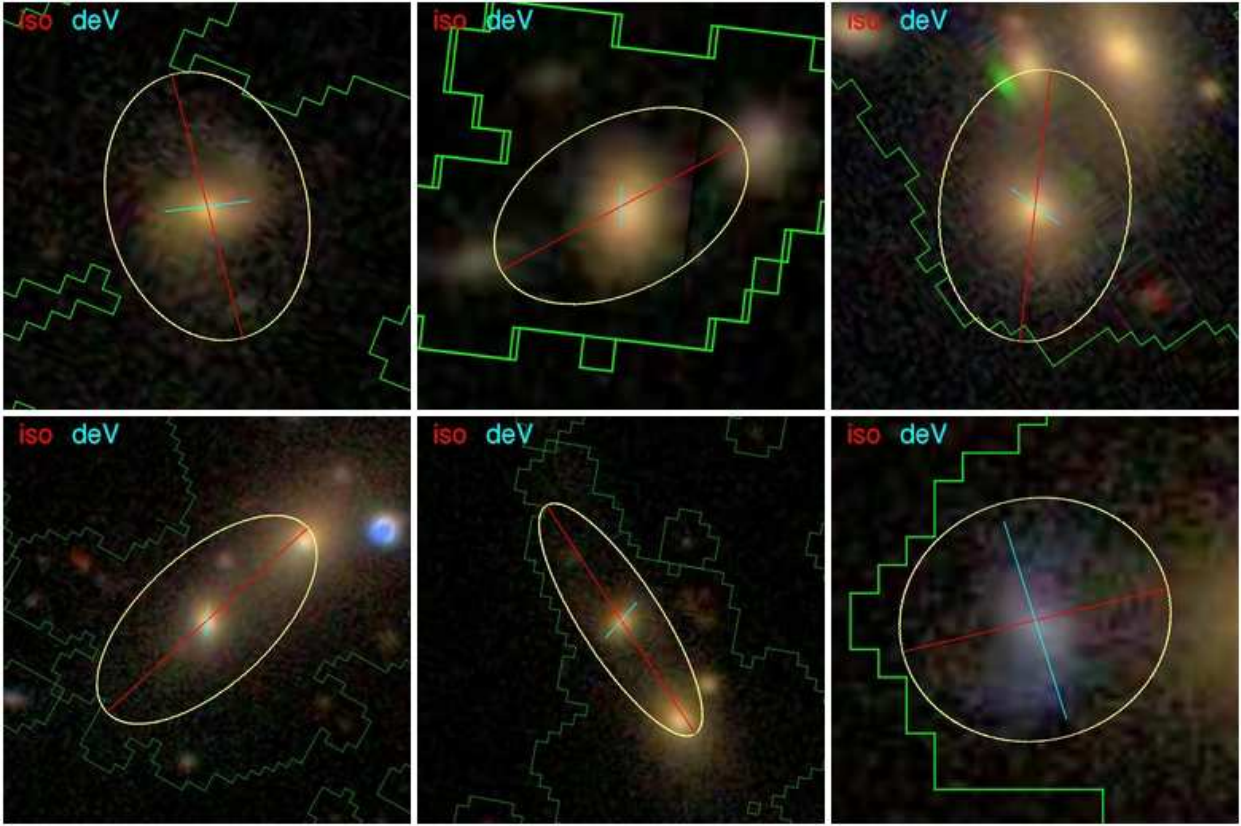}
\caption{\label{fig:PA_FakeDiff} Position angles measured with isophotes and
galaxy models can provide highly discrepant results. Unlike the previous
example (Figure \ref{fig:PA_RealDiff}) where two discrepant position angles
both correspond to obvious galactic features, the objects above have nearby
neighbors which distort the appearance of the isophotes nearby. In several of
the cases above, this systematic effect could easily mimic a hostward
elongation signal. } 
\end{figure}

Figure \ref{fig:PA_FakeDiff} shows illustrative examples of isophotal
contamination. In a number of cases, failure to separate the light from other
objects results in an isophotal PA which does not physically reflect the system
in question. Although not as prevalent as isophotal twisting, such cases are
responsible for a significant fraction of discrepant PA measurements.

Importantly, we note that we do not expect random (i.e., isotropic) PA
contamination for galaxies in groups and clusters. In such environments, galaxy
spatial density is a rapidly declining function of separation from the center
($\DR$). As a result, a given group or cluster member is most likely to have an
optical companion radially interior to its location. We thus expect that
isophotal contamination due to chance juxtaposition of galaxies will
preferentially skew isophotal PA in the radial (hostward) direction. The degree
to which this occurs will depend strongly on the how rapidly galaxy number
density falls off with increasing separation from the cluster center.

From this investigation we conclude that care is required to ensure that only
accurate PA measurements are included in alignment analysis. This is
particularly important for contamination cases where the best-fit isophotal
ellipse is highly elongated, giving the illusion of high accuracy. Elimination
of discrepant cases has a profound effect on the measured hostward alignment
signal. We believe the use of different PA measurements may be responsible for
several conflicting previous results (see discussion in \S\ref{ss:comp_elong}).


\subsection{Spectroscopic Survey Limitations} \label{ss:survey_limits}

In some cases, the limitations of the SDSS spectroscopic sample restrict our
ability to properly characterize alignment. Fiber collisions make it difficult
to observe galaxies with small projected separations. On SDSS plates, the
spectroscopic fibers (3'' each) can be set no closer than 55'' apart. Although
this limit is mitigated somewhat by plate tiling, many of the closest neighbors
in projection will be absent. In addition, the finite (640) number of fibers
allowed per plate ultimately limits the fraction of galaxies that can be
observed in dense fields. As a result, the completeness of a galaxy group
generally decreases with increasing group size. The inability to fully sample
the innermost regions of groups and clusters could systematically affect the
measured alignment signal. Of importance to our exploration of the Holmberg
Effect in the SDSS spectroscopic sample, some bright nearby neighbors may be
absent because of these effects. If such omissions were common, we would
incorrectly identify many objects as isolated galaxies. Whether or not the
robust elimination of interlopers afforded by spectroscopic redshifts outweighs
the potential statistical benefits of the larger photometric sample for
detecting alignment trends is a difficult question which we do not attempt to
answer here. It is likely, however, that exclusive use of galaxies from the
spectroscopic sample will not provide the most accurate picture of galaxy
alignment at separations approaching the fiber collision limit.

The magnitude limit of the spectroscopic survey may also affect our results. It
was observed in early cluster studies \citep[see
][]{Sastry_68,noonan72,austin_peach_74} that the apparent shape of a cluster
(isopleths) may vary as a function of the galaxy flux limit, so that the
distribution of the brightest members may not be representative of the overall
population. A similar effect due to either the spectroscopic magnitude limit or
fiber collisions is possible with these data. 

The survey magnitude limit also likely hinders our ability to properly
reproduce the original selection criteria of \citet{holmberg_69} and robustly
assess the existence of the Holmberg Effect. \citet{holmberg_69} mandated a
host-satellite mass ratio of at least 25 (corresponding to a flux ratio of
roughly 3.5 magnitudes). With the limiting SDSS survey magnitude of 17.77, only
host galaxies of $r \lesssim 14$ are bright enough to have
spectroscopically-identified ``Holmberg-like'' companions. Of the 572,495
galaxies in our dataset, only 2,775 are this bright. We find only 1178 such HGs
in our group catalog.

To make matters worse, scattered light in the SDSS telescope increases sky
brightness around bright galaxies \citep{DR6}. This effect is believed to
conceal fainter nearby galaxies, resulting in an observed dearth of faint
galaxies near bright galaxies \citep{DR6}. Several recent studies of weak
lensing and SG anisotropy mask objects at very close separations to avoid these
potential selection biases (for example, objects in the innermost $30\hkpc$ and
$35\hkpc$ are excluded in \citealt{mandelbaum_2005} and \citealt{bailin_2007},
respectively). In particular, this seriously complicates application of
isolation criteria.

\subsection{Isolation}

Well-selected isolation criteria are of equal importance to major axis
preference and to hostward elongation, both of which have now been observed to
varying degrees in multiple independent investigations. Several of these
studies \citep[e.g.,][]{brainerd_2005_ani,AB_2006a,yang_2006} employ
aggressive selection criteria in order to extract a sample of isolated objects.
Among these isolated galaxy systems, they observe strong SG distribution
anisotropy favoring the HG major axis. The aforementioned studies additionally
report near-perfect SG isotropy around blue hosts (i.e., no sign of the
Holmberg Effect).

Recently, \citet{bailin_2008} performed a systematic evaluation of the various
selection criteria used in recent alignment studies. They conclude, based on
examination of mock catalogs and simulations, that most previous efforts to
select only isolated galaxy systems were ineffective. Objects people believed
were truly isolated systems were statistically shown to include many group and
cluster members. In reality, the SDSS spectroscopic survey alone does not, in
general, include all potential neighbor galaxies, which makes application of
isolation criteria very difficult. Based on our satellite anisotropy results
(\S\ref{ss:results_map}) and the HG and SG properties most closely associated
with the alignment signal, we conclude that the alignment signal is largely due
to the contribution of group and cluster members despite the ostensibly
isolated nature of our group samples.


\section{Previous Detections and Comparisons} \label{s:comparison}


\subsection{Satellite Galaxy Anisotropy} \label{ss:comp_map}

The tendency for satellite galaxies to have anisotropic distributions favoring
the major axis of their host galaxies has been observed by many authors using
SDSS data \citep{brainerd_2005_ani,yang_2006,azzaro_2007,falten1,bailin_2008}.
The enhancements of this signal among red HG-SG pairs
\citep{yang_2006,azzaro_2007,bailin_2008} and with decreasing radial separation
\citep{yang_2006} have been independently observed. Our results are consistent
with these other recent findings in both in strength and nature with these
other recent findings.

We build on these results by demonstrating the monotonic dependence of major
axis preference on HG and SG light profile concentration, physical separation
(both $\DV$ and $\DR$) and group size. In \S\ref{ss:results_map} we show that
red, highly concentrated SGs of red, highly concentrated HGs preferentially
reside near the projected major axis of their hosts. We also show that this
effect increases in strength with decreasing physical separation. Lastly, but
equally importantly, we observe that the alignment strength increases with
increasing group size.

Galaxy clusters are known to produce a similar effect. Many past investigations
have discovered elongated galaxy cluster isopleths which align strongly with
the major axis of the brightest cluster galaxy (BCG) \citep[e.g.\
][]{carter_metcalfe_1980,binggeli_1982}. Indeed, BCG - cluster alignment is
sufficiently strong that the BCG PA can be used as a proxy for the PA of the
cluster galaxy distribution. 

As would be expected in cluster environments, the galaxies which give rise to
our measured anisotropy are mainly red, highly concentrated ellipticals. In the
literature we find good agreement between our alignment findings and those from
deep campaigns of individual galaxy clusters (e.g., the histograms of
\citealt{carter_metcalfe_1980} and \citealt{binggeli_1982}). This further
supports our suspicion that group and cluster members produce the majority, if
not all, of the observed anisotropy signal.


\subsection{Hostward Elongation} \label{ss:comp_elong}

We observe hostward elongation in our SG sample (\S\ref{ss:res_elong}) at the
few percent level (measured in fractional galaxy density change), in fair
agreement with numerous previous studies. We find that the alignment signal is
strongest among high $\FD$ ($\FD>0.9$) SGs irrespective of host type. We also
observe an enhancement among red SGs (independent of HG color). Most
importantly, we report a connection between the type of PA used and the
apparent alignment strength which may explain the variation among several
previous results.  

For some time, authors have suspected that intrinsic galaxy alignment effects
may act as a contaminant in weak lensing surveys \citep{croft_metzler_2000}.
\citet{lee_pen_2001} predicted that intrinsic alignments would dominate the
weak lensing signal for spiral galaxies in SDSS. \citet{bernberg_2002} found no
significant tangential elongation in a study of 1819 galaxies from the 2dFGRS,
limiting the contamination to $\sim 20\%$. \citet{hirata_2004} found no
evidence for intrinsic alignment of satellite galaxies measured with the
adaptive moments technique \citep{bernjar}. \citet{mandelbaum_2005} also
concluded that contamination due to intrinsic galaxy alignment is minimal.

Using our entire SG sample, we find that the frequency of SGs oriented radially
towards their hosts is 6\% greater than that of SGs oriented perpendicular to
their hosts (\S\ref{ss:res_elong}). This anisotropy is strongest among
high-$\FD$ SGs, for which the fractional change steepens to $-14\%$.  These
findings are generally consistent with those mentioned above. On the other
hand, \citet{pereira_kuhn_2005} report a much stronger radial alignment in SDSS
data with a sample of 85 X-ray selected clusters. They observe a $-25.5\%$
change at $4\sigma$ significance. \citet{AB_2006a} also find a much stronger
radial elongation effect. They find that SGs located at small projected
separation ($\DR<100\hkpc$) from relatively isolated hosts exhibit a
$\sim-20\%$ density change between radial and tangential orientation.
\citet{falten1} reports a similarly strong effect for red satellites at low
projected separation. 

The last three studies, mentioned above, relied exclusively on isophotal PA
measurements, which, we believe, are suspect. We demonstrated the impact of PA
type (isophotal or model-based) on the measured hostward elongation signal in
\S\ref{ss:discrepant_PA}. By varying our PA selection, we can recover results
from a wide range of recent investigations. We conclude specifically that
relying solely on SDSS isophotal PA produces a stronger radial (hostward)
elongation signal than does the exclusive use of matched models or adaptive
moments. We present our interpretations and further discussion in
\S\ref{s:discussion}.


\section{In Search of the Holmberg Effect} \label{s:holmberg}

\citet{holmberg_69} observed that the distribution of nearby (projected radial
separation $<40\hkpc$) satellite galaxies (SGs) around inclined, isolated
spiral host galaxies (HGs) was heavily concentrated near the HG projected minor
axis. In fact, after statistically subtracting the expected
isotropiccally-distributed field galaxies, Holmberg found effectively no SGs
within $30\DEG$ of the projected HG major axis ($\theta<30\DEG$). Since the
host galaxies were inclined spirals, this suggested that SGs of such hosts have
primarily polar orbits.

\citet{holmberg_69} selected isolated systems for several reasons. First, such
systems are dynamically simple in comparison to larger groups. In addition,
stringent isolation criteria could be used to decrease the frequency of
interlopers in the SG sample. To evaluate a galaxy's degree of isolation,
\citet{holmberg_69} first defined a circular survey region of $40\hkpc$
projected radius around each HG.  Next, he estimated the (stellar) mass of each
HG and SG companion from luminosity and morphological type. Isolation required
that all SGs within twice the survey radius ($80\hkpc$) have less than $1/5$
the mass of the HG and all SGs within the survey radius ($40\hkpc$) have less
than $1/25$ the HG mass.  Systems which passed those criteria and were
sufficiently inclined ($q\le0.53$) were included in the analysis.

In practice, exactly replicating these conditions using the SDSS spectroscopic
galaxy sample is essentially impossible, primarily because of spatial sampling
constraints (i.e., ``fiber collisions,'' see \S\ref{ss:survey_limits}).
Insufficient spatial sampling causes multiple problems. Firstly, the very small
typical separations of Holmberg's original SG sample ($40\hkpc$ is roughly
$4\times$ the HG semi-major axis length) greatly intensifies the incompleteness
of the sample due to the fiber collision effect. After removing likely blends
and misidentifications (SGs with $\DR < \prad$ or $\DR < iso_R$) we find only
132 SGs around 130 inclined ($\qiso < 0.53$ and $\qmom < 0.53$) HGs in our
original group sample (i.e.\ without any isolation criteria applied) with $\DR
< 40\hkpc$. Without the ability to identify all nearby galaxies of comparable
size and brightness, it is impossible to properly apply Holmberg's isolation
criteria to our galaxy sample.


\subsection{An Isolated Sample}

We first attempt to apply Holmberg's selection criterion (approximately) to our
group sample in order to demonstrate the difficulties with reproducing his
analysis precisely with the SDSS spectroscopic sample. Unable to exactly
reproduce his galaxy mass estimates, we apply the same cuts using bolometric
luminosity (calculated with the LRT codes). The modified group selection
procedure incorporates the following criteria: 
\begin{eqnarray} 
&&(\lbol)_{\mHG} \ge 25 \, (\lbol)_{\mSG} \; \mathrm{for} \; \DR \le 40\hkpc \\ 
&&(\lbol)_{\mHG} \ge \;5\, (\lbol)_{\mSG} \; \mathrm{for} \; \DR \le 80\hkpc. 
\end{eqnarray} 
All hosts which fail these tests are removed from the HG list. The remaining
groups comprise our `isolated' sample. We identify inclined HGs by axis ratio,
requiring that $\qmom < 0.53$ and $\qiso < 0.53$. Unfortunately, this procedure
yields only 2 SGs of 2 HGs within the neighbor volume employed by
\citet{holmberg_69}.


\subsection{More Distant Companions: a Marginal Effect}
\label{ss:marginal_holmberg}

Unable to replicate the analysis of \citet{holmberg_69} with SDSS spectroscopic
data, we instead proceed with an examination of more distant SG companions. We
select SGs of inclined isolated systems as above but relax the radial
separation constraint, including SGs as distant as $500\hkpc$. In the spirit of
selecting relatively isolated HGs, we exclude SGs from systems with more than
25 detected satellites within $1\hmpc$. From among these, we select the SGs
with $15\times\SG\;\lbol<\HG\,\lbol$ and $\SG\,\FD<0.10$. These last few
criteria specifically pick out smaller SGs while avoiding clusters and big
groups.

\begin{figure} 
\epsscale{0.95}
\plotone{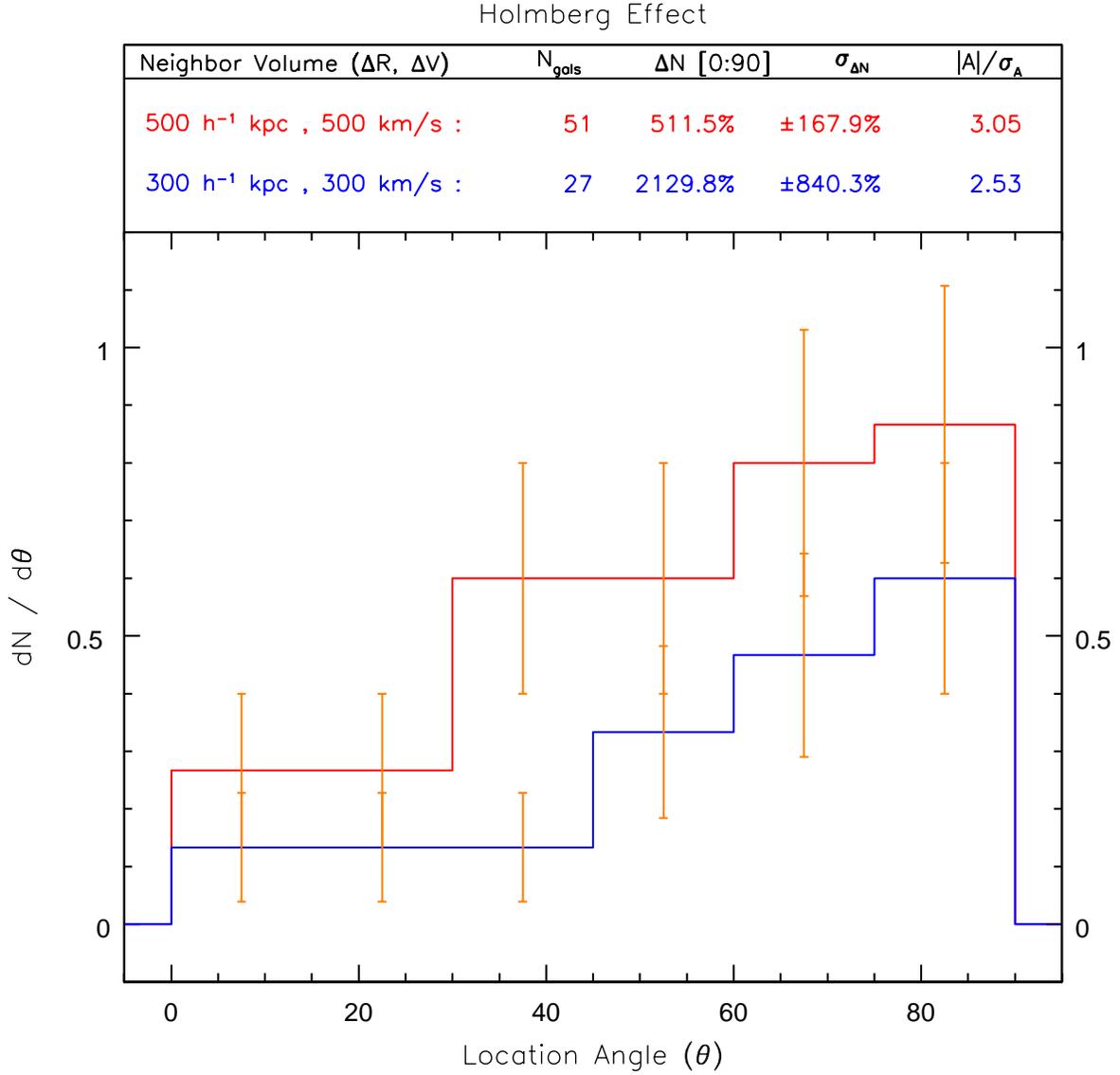}
\epsscale{1.00}
\caption{\label{f:holmberg2} Holmberg-like effect seen among select satellite
galaxies (SGs). In the above histograms we include only groups with inclined
($\qmom < 0.53$ and $\qiso < 0.53$), blue host galaxies (HGs) with fewer than
25 identified satellites within $1\hmpc$. We further require that isophotal and
model HG position angle (PA) agree well ($\DPA < 15\DEG$). Within these
systems, we keep SGs that are $15\times$ less luminous than their hosts and
satisfy $|\DV| < 500\kms$. We eliminate all SGs within 5 Petrosian radii of the
host to exclude mis-detected bright knots in the host galaxy. The red curve
above includes objects with $\DR<500\hkpc$ and $\DV<500\kms$ while the blue
curve extends only to $\DR<300\hkpc$ and $\DV<300\kms$. }
\end{figure}

Applying the above criteria, we observe a preference for the minor axis among
the remaining SGs (Fig.\ \ref{f:holmberg2}). Among SGs with $\DR<500\hkpc$ and
$\DV<500\kms$ (51 galaxies), we find few galaxies (4 galaxies per $15\DEG$ bin)
near the HG major axis ($\theta<30\DEG$). The density increases to 9 galaxies
per bin for $30\DEG<\theta<60\DEG$ and to 12 and 13 galaxies per bin for
$60\DEG<\theta<75\DEG$ and $75\DEG<\theta<90\DEG$, respectively. This roughly
3-fold increase indicates anisotropy at $3\sigma$ significance. For SGs with
$\DR<300\hkpc$ and $\DV<300\kms$ (27 galaxies), we find a low, constant galaxy
density (2 galaxies per $15\DEG$ bin) for $\theta<45\DEG$. For $\theta>45\DEG$,
we find that galaxy density rises steeply, such that galaxies with
$75\DEG<\theta<90\DEG$ are 4.5 times as numerous (9 galaxies) as those with
$\theta<45\DEG$. For the latter case, although the fractional slope is quite
large, the dwindling galaxy counts ultimately reduce the significance of the
deviation from isotropy to approximately $2.5\sigma$. Keep in mind that rather
than including all of the satellites in the galaxy systems, we have selectively
included and excluded SGs with different properties to obtain this behavior. We
caution that that our results depend quite sensitively on the specific
selection criteria we use. With so few objects, further subdivision is not
practical. 

In summary, we observe that a subset of SGs with certain properties favor the
HG minor axis within $500\hkpc$ projected radius at marginal significance. The
strength of this trend may increase with decreasing $\DR$, but the accompanying
decrease in the total number of SGs decreases the significance of this result.
Whether or not this apparent anisotropy and preference for the HG minor axis is
due entirely to selection effects remains to be seen.


\subsection{Difficulties and Discussion} \label{ss:comp_holmberg}

Although we observe a trend that resembles the Holmberg Effect, we are unable
to confirm or refute the findings of \citet{holmberg_69} in the present work.
\citet{holmberg_69} enforced isolation criteria using a mass cut. In practice,
his mass determination is difficult to reproduce. As a substitute, we impose
the same cuts in bolometric luminosity ($\lbol$ computed with LRT). We also use
SDSS $\FD$ and K-corrected galaxy colors in lieu of confirmed morphological
type.

A more important difference is the spatial extent of satellite survey regions.
The \citet{holmberg_69} sample is restricted to faint nearby (typically within
4 galaxy radii) companions. Very few such galaxies even exist in the SDSS
spectroscopic sample \citep[only find 2 after imposing criteria
of][]{holmberg_69}. At close separations, \citet{holmberg_69} requires a
host-satellite mass ratio of at least 25. In practice, many such galaxies will
be near the magnitude limit of the spectroscopic survey. Those that are
sufficiently bright may be undersampled because of fiber collisions (see
\S\ref{ss:survey_limits}).

\citet{zaritsky_97_ani} detect a Homberg-like SG anisotropy using their own
independent sample \citep{zaritsky_97_catalog} at larger radii ($\ge
200\,\kpc$). They find inconsistency with isotropy at the 99\% confidence
level, reporting that 72 of their 115 SGs (63\%) reside at $\theta > 45\DEG$.
This corresponds to a 217\% fractional increase in the number of SGs aligned
with the minor axis relative to the major axis, assuming the underlying
distribution is linear in $\theta$.

We find an alignment signal resembling that of \citet{holmberg_69}, albeit at
larger projected separations. In magnitude and radial extent, our results are
generally consistent with those of \citet{zaritsky_97_ani}, although at
marginal significance. We further caution that our findings are highly
sensitive to selection criteria and may be the result of selection effects. 


\section{Discussion and Conclusions} \label{s:discussion}

We conclusively observe an anisotropy in the distribution of satellite galaxies
(SGs) in systems with red, highly concentrated (chiefly elliptical) host
galaxies (HGs). Within these systems, the alignment trend strength is further
enhanced by nearly a factor of two for red, high-concentration SGs (also mainly
ellipticals). We find conclusive evidence of increasing alignment strength with
decreasing physical separation (both in projected radius $\DR$ and velocity
separation $\DV$), noting that the fractional anisotropy among the closest
satellites ($\DR<250\hkpc$) is 40\% greater than that seen among more distant
($500\hkpc<\DR<1000\hkpc$) companions. In addition, the strength of the
alignment trend increases with spectroscopically-identified group size (see
\S\ref{ss:results_map}).

Although these results agree well with several other recent studies (see
\S\ref{ss:comp_map}), we offer a slightly different interpretation. Based on
the nature of the galaxies which give rise to this effect, we suspect that
members of large groups and clusters contribute significantly to, or even
dominate, this alignment trend. The well-known alignment of the members of
galaxy clusters with the orientation of the brightest member has a comparable
magnitude and sign. A recent thorough inspection of popular selection criteria
\citep{bailin_2008} determined that extremely stringent criteria are required
to generate a sample of truly isolated hosts and satellites. They find that
with less stringent criteria, group and cluster members will be ubiquitous in
ostensibly isolated SDSS galaxy samples.

The cluster origin offers a natural explanation for this alignment effect.
Cluster shape is defined by the spatial distribution of member galaxies. These
members form in the high-density filaments, creating an anisotropy in the
distribution of satellites during the formation process. Although dynamical
processes may alter the distribution somewhat as time passes, numerical work
indicates that the cluster shape will retain the memory of the initial member
galaxy distribution (see e.g., \citealt{knebe_2004} and \citealt{libesk_2005}).
The brightest cluster galaxies (BCGs) could acquire their alignment with this
anisotropic distribution of member satellites through tidal forces exerted by
the large-scale structure during their formation. Alternatively, mergers and
accretion events drawing from the already-anisotropic SG distribution could
create this alignment. See \citet{bailin_2008} for further details and
discussion of this scenario.

We also find significant evidence of hostward elongation among satellite
galaxies (SGs). Although this effect has been observed before, we are able to
reconcile many previously conflicting results. Among the attributes we
investigate, hostward elongation depends most strongly on SG concentration and
color. We find effectively no significant direct dependence on HG properties.
Ultimately we conclude that the hostward elongation is real, but not nearly as
strong an effect as the major axis anisotropy mentioned above. 

More importantly, we find that the detection of hostward elongation is a strong
function of the method used to measure and define position angle of the galaxy.
We find that, at low projected separation ($\DR<100\hkpc$), the radial
alignment signal calculated with isophotal PA sharply increases in strength. We
observe no such increase among \textit{the same galaxies} using de Vaucouleurs
PA. We further find that excluding objects whose PA measurements differ by more
than $15\DEG$ eliminates this discrepancy. This explains why some authors
(e.g., \citealt{pereira_kuhn_2005}, \citealt{AB_2006a}, \& \citealt{falten1}),
who relied solely on isophotal PA for galaxy orientation, observed a stronger
radial alignment effect than those who used galaxy moments (e.g.,
\citealt{bernberg_2002}, \citealt{hirata_2004}, \& \citealt{mandelbaum_2005}).
Given that PA determination plays such a crucial role in galaxy alignment
studies, we stress the prevalence of these discrepant PAs and their significant
potential impact.

At the same time, this observed discrepancy raises new questions. Host galaxies
(HGs) in our group sample appear immune to this phenomenon
(\S\ref{ss:discrepant_PA}). The rate at which SGs are affected by this appears
to increase dramatically at small projected separations from the HG. Observing
the anomalous objects by eye, we find that most exhibit strong isophotal
twisting (e.g.\ barred spiral galaxies), although there is a significant
(nearly 1/3) contribution from isophotal contamination. As outlined previously,
isophotal contamination may induce a radial alignment bias if spatial galaxy
density is a decreasing function of projected separation.

For cases of severe isophotal twisting, position angle is an ill-defined
quantity. Since the mechanisms of galaxy production and alignment within their
haloes is still highly uncertain, it is not clear how to best incorporate
galaxies with extreme isophotal twisting into an alignment investigation such
as this. For the same reason, we are unsure how to interpret a radial
elongation tendency. Some numerical simulations \citep[e.g.][]{falten2} show
a strong tidally-induced radial alignment effect in dark matter haloes.
Whether or not such orientations would also apply to the luminous regions of
galaxies within these haloes is unclear.

Regardless of the cause, measuring the intrinsic radial elongation signal of
galaxy systems is an important step in the application of weak lensing to
precision cosmology. In general agreement with other authors \citep[e.g.\
][]{bernberg_2002,mandelbaum_2005}, we believe that a radial elongation signal
with the magnitude we have found in SDSS spectroscopic galaxy sample will
minimally influence galaxy-galaxy lensing at the levels of precision currently
required.  Further investigation, however, will be required to determine
whether or not this hostward elongation is observed among fainter galaxy
populations (i.e., in the SDSS photometric galaxy sample).

A third alignment trend, the Holmberg Effect, remains improperly tested. The
magnitude and spatial sampling limitations of the SDSS spectroscopic survey
combined with systematic effects in the vicinity of bright galaxies effectively
preclude implementation of Holmberg's original selection criteria. Instead, we
explore the distribution of SGs at larger projected separation. We find that,
although the whole SG population around inclined, isolated spirals shows no
detectable anisotropy, specific SG subsets do. SGs from smaller systems ($\NSG
< 25$) with $\FD < 0.15$ and $\lbol < 1/15$ that of their host overwhelmingly
prefer the HG minor axis, albeit at marginal ($3.4\sigma$) significance owing
to the relatively small number of objects that satisfy these criteria.

Through improved convergence of independent observational results, we have
uncovered and characterized several different galaxy alignment trends. However,
further refinement is required before high-accuracy alignment prescriptions are
available for weak lensing and simulations of galaxy formation and evolution.
Importantly, much of this remaining work pertains to identification and removal
of systematic effects such as position angle uncertainty that currently prevent
a more thorough understanding of intrinsic galaxy alignment effects. 


\acknowledgements

We would like to thank Roberto Assef for his assistance in adapting his
low-resolution template code to our data set. We would also like to thank Oleg
Gnedin for his useful comments and suggestions.

Funding for the SDSS and SDSS-II has been provided by the Alfred P. Sloan
Foundation, the Participating Institutions, the National Science Foundation,
the U.S. Department of Energy, the National Aeronautics and Space
Administration, the Japanese Monbukagakusho, the Max Planck Society, and the
Higher Education Funding Council for England. The SDSS Web Site is
http://www.sdss.org/.

The SDSS is managed by the Astrophysical Research Consortium for the
Participating Institutions. The Participating Institutions are the American
Museum of Natural History, Astrophysical Institute Potsdam, University of
Basel, University of Cambridge, Case Western Reserve University, University of
Chicago, Drexel University, Fermilab, the Institute for Advanced Study, the
Japan Participation Group, Johns Hopkins University, the Joint Institute for
Nuclear Astrophysics, the Kavli Institute for Particle Astrophysics and
Cosmology, the Korean Scientist Group, the Chinese Academy of Sciences
(LAMOST), Los Alamos National Laboratory, the Max-Planck-Institute for
Astronomy (MPIA), the Max-Planck-Institute for Astrophysics (MPA), New Mexico
State University, Ohio State University, University of Pittsburgh, University
of Portsmouth, Princeton University, the United States Naval Observatory, and
the University of Washington.


\clearpage


\appendix
\singlespace
\begin{small}

\section{Expected Signal-to-Noise} \label{a:SN_derive}

\newcommand{\binsum}{\sum_{i=1}^{n_b}}

Here we derive the expected signal-to-noise ratio (S/N) of a trend detection
using our histogram-fitting method (\S \ref{ss:significance}). Here S/N is the
square root of the $\Delta \chi^2$ improvement of a weighted linear
least-squares fit relative to a constant value (mean). Larger $\Delta \chi^2$
values constitute stronger evidence for anisotropy. We define:
\begin{equation} S/N = \sqrt{|\Delta \chi^2|}.  \end{equation}

In our linear model, the histogram has $n_b$ bins in the angle $\theta$. Each
bin $i$ contains 
\begin{equation}
%
N_i = A \, \theta_i \, d\theta + \overline{N}
\end{equation} 
galaxies, where $\overline{N} = N_{tot}/n_b$ is the mean number of galaxies per
bin, $d\theta$ is the bin width, $N_{tot} = \binsum N_i$ is the total number of
galaxies in the sample, and $A$ is the best-fit slope of a linear fit to the
histogram bins. The constant model assumes that $N_i = \overline{N}$.

The improvement of a linear least-squares fit over the constant value is the
squared difference between the two models divided by the individual bin counts,
summed over all histogram bins. The $\overline{N}$ terms in the linear and
constant models cancel, leaving:
\begin{equation} 
\Delta \chi^2 = \binsum \frac{(A \, \theta_i \, d\theta)^2}{N_i}.
\end{equation}

To simplify the calculation, we assume that the $N_i$ are roughly equal and
that $N_i \approx \overline{N}$. Removing bin-independent terms from the
summation, we have:
\begin{equation}
\Delta \chi^2 = \frac{A^2}{\overline{N}}\,d\theta \,\binsum \theta_i^2 d\theta.
\end{equation}

To simplify the calculation, we define $\Delta \theta$ to be half the histogram
width ($\Delta \theta = 45\DEG$ in this case). We can relate this to bin width
($d\theta$) by noting that $2\Delta \theta = n_b \, d\theta$. Substituting
this outside the summation, we have:
\begin{equation}
\Delta \chi^2 = \frac{A^2}{\overline{N}} \, \frac{2 \Delta \theta}{n_b} \,
\binsum \theta^2 d\theta.
\end{equation}

We convert this to an integral in the limit $n_b \rightarrow \infty$, leaving
\begin{equation}
\Delta \chi^2 = \frac{A^2}{\overline{N}} \, \frac{2\Delta\theta}{n_b} 
\int\limits_{-\Delta\theta}^{\Delta\theta} \theta^2 d\theta = 
\frac{A^2}{\overline{N}} \, \frac{2\Delta\theta}{n_b} \, \frac{2}{3} \, 
\Delta \theta^3 = \frac{A^2}{\overline{N}} \, \frac{4\Delta\theta^4}{3n_b} = 
\frac{4 A^2 \Delta\theta^4}{3N_{tot}}.
\end{equation}

Finally we have,
\begin{equation}
\frac{S}{N} = \sqrt{|\Delta \chi^2|} = \sqrt{\frac{4}{3}} \, 
\frac{A\Delta\theta^2}{\sqrt{N_{tot}}}.
\end{equation}

\end{small}

\clearpage


\clearpage

\newcommand{\LAInset}{The inset directly compares the normalized fractional
slopes of the different subsamples. $N / N_0$ is normalized galaxy density,
where $N_0$ is the galaxy density along the host galaxy projected
major axis $N(\theta=0)$. The dashed black line indicates zero slope
(isotropy).}

%
\begin{figure}
%
\plotone{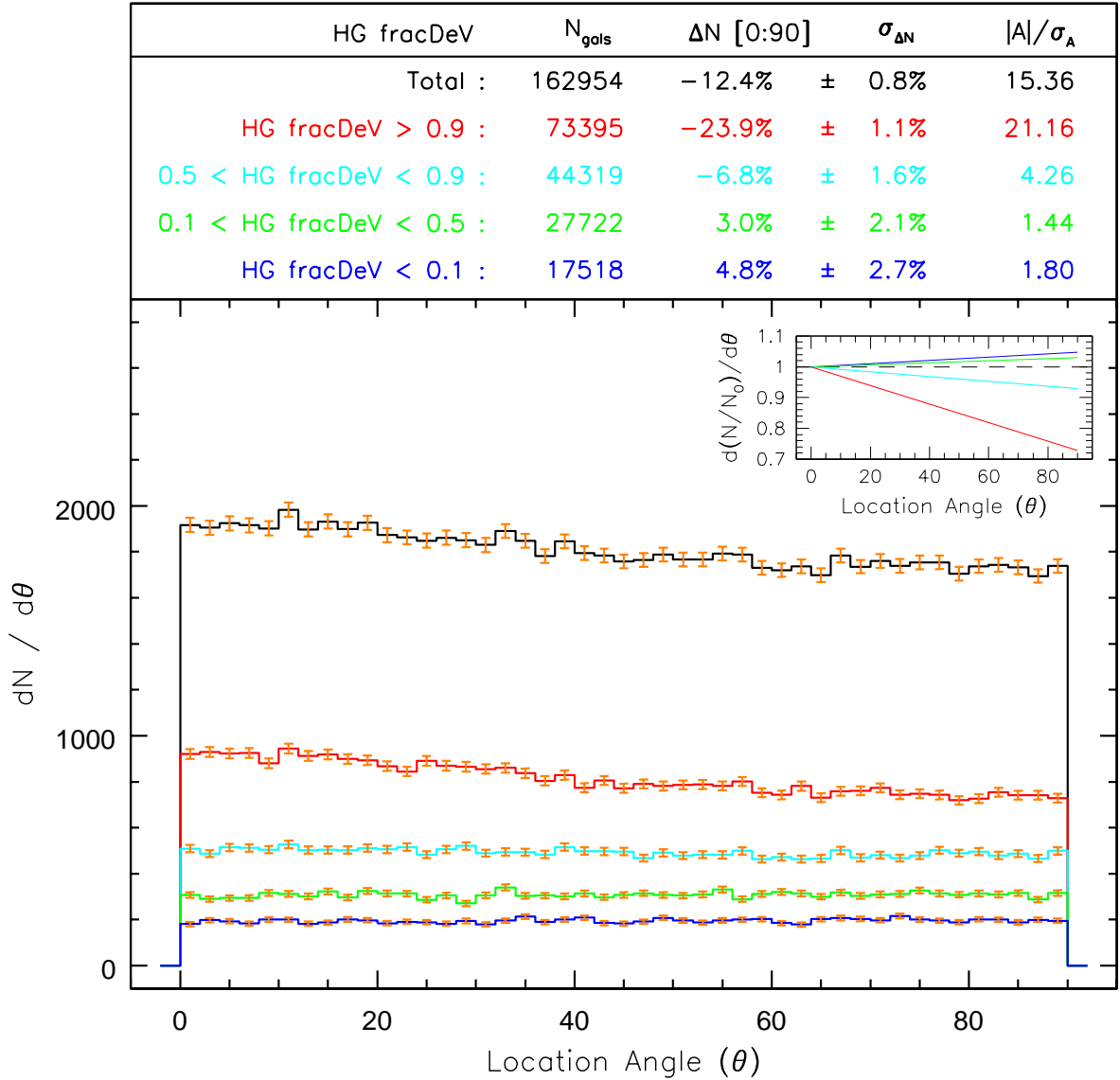}
\caption{\label{f:d1_hg_fdev} Satellite location angle ($\theta$) for different
subsamples of host galaxy $\FD$ with $1\sigma$ Poisson error bars. The upper
panel provides the following subsample-specific data: the number of galaxies
($N_{gals}$), the fractional slope as a percent decrease ($\Delta N [0:90]$),
uncertainty in the percentage slope ($\sigma_{\Delta N}$), and anisotropy
significance ($|A|/\sigma_A$). \LAInset }
\end{figure}

\begin{figure}
%
\plotone{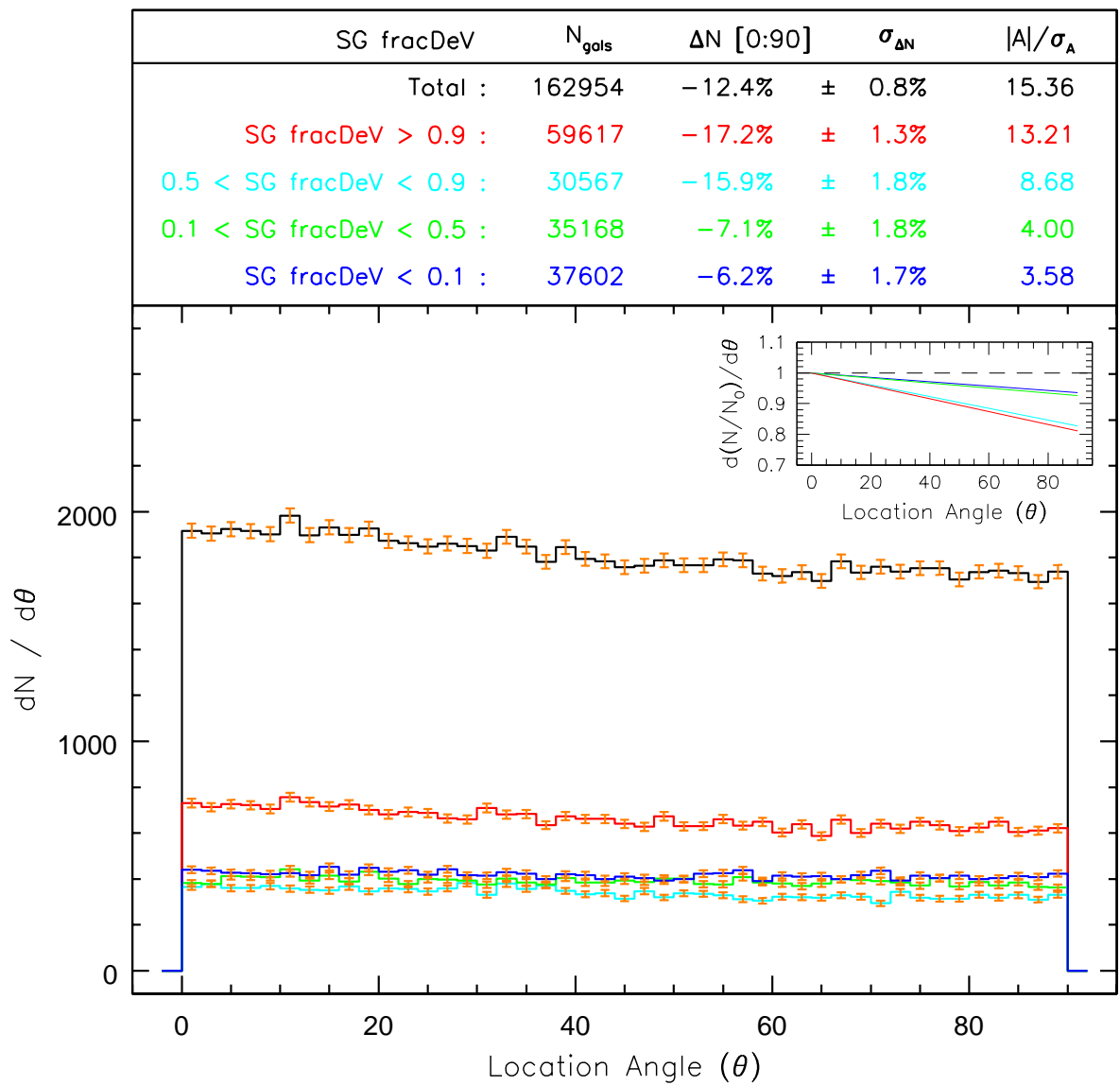}
\caption{\label{f:d1_sg_fdev} Satellite location angle ($\theta$) for different
subsamples of satellite galaxy $\FD$ (concentration). The format of this figure
is identical to that of Figure \ref{f:d1_hg_fdev}. All four subsamples show
some preference for the host major axis. de Vaucouleurs profile-dominated
satellite galaxies ($\FD>0.5$) prefer the major axis most strongly. }
\end{figure}

\begin{figure}
%
\plotone{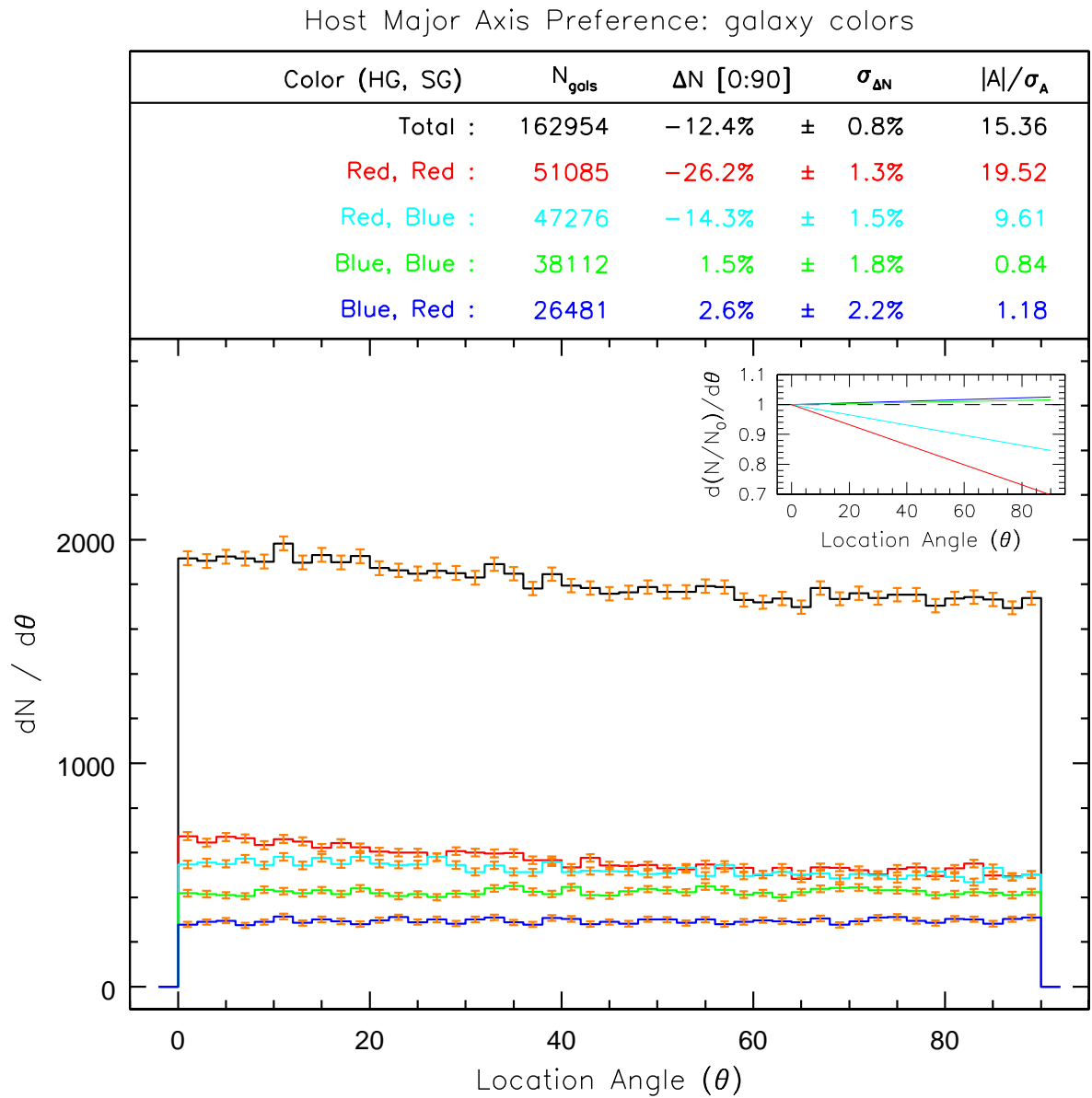}
\caption{\label{f:d1_color} Satellite location angle ($\theta$) as a function
of host galaxy (HG) and satellite galaxy (SG) colors. The format of this figure
is identical to that of Figure \ref{f:d1_hg_fdev}. We perform color separation
using the division of \citet{bailin_2007} (Fig.\ \ref{f:galCMD}). Major axis
preference is restricted to satellite galaxies of red hosts and is strongest
specifically for red satellites of red hosts. Satellites of blue host galaxies
are generally consistent with isotropy. } 
\end{figure}

\clearpage

\begin{figure}
%
\plotone{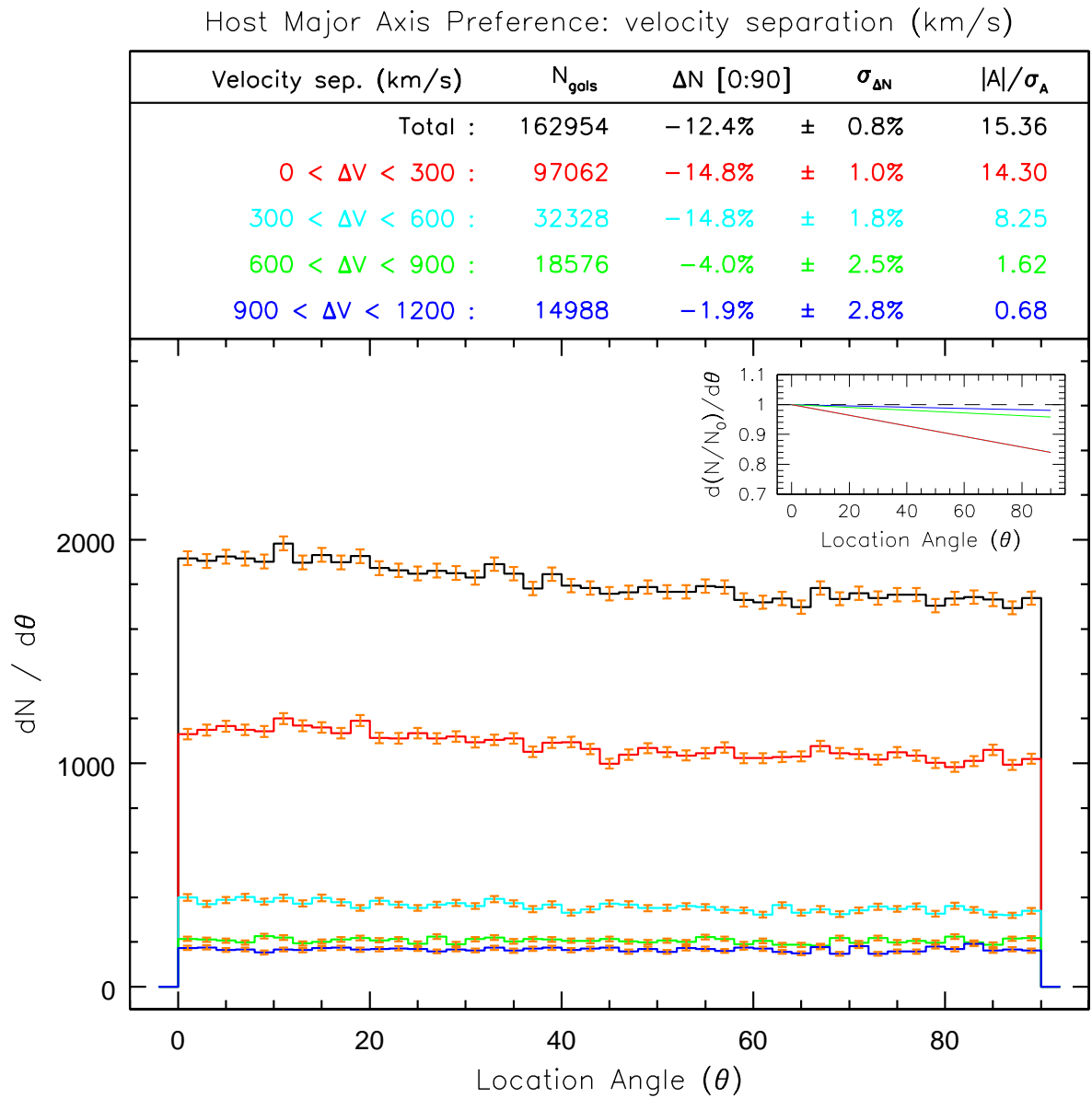}
\caption{\label{f:d1_deltaV} Satellite location angle ($\theta$) for different
subsamples of host-satellite velocity separation ($\DV$). The format of this
figure is identical to that of Figure \ref{f:d1_hg_fdev}. Major axis preference
is restricted to the two lower $\DV$ bins ($\DV<600\kms$). At large $\DV$
($>900\kms$), the satellite galaxy distribution is consistent with isotropy. } 
\end{figure}

\begin{figure}
%
\plotone{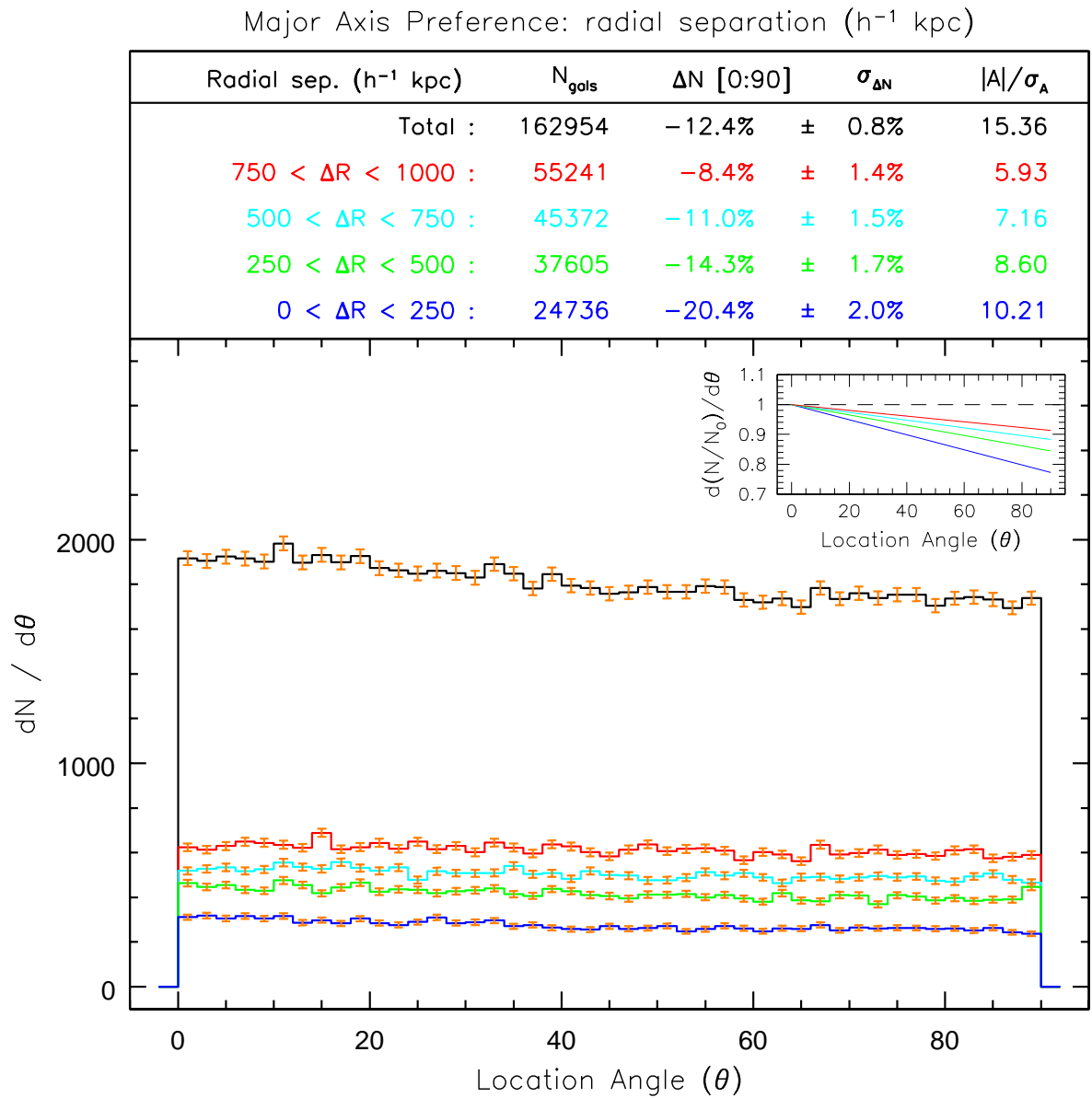}
\caption{\label{f:d1_radius} Satellite location angle ($\theta$) for different
subsamples of projected host-satellite radial separation ($\DR$). The format of
this figure is identical to that of Figure \ref{f:d1_hg_fdev}. Major axis
preference is strongest and most significant at close separations
($\DR<250\hkpc$). Despite increasing subsample size, this anisotropy decreases
steadily in strength and significance with increasing radial separation. } 
\end{figure}

\begin{figure}
%
\plotone{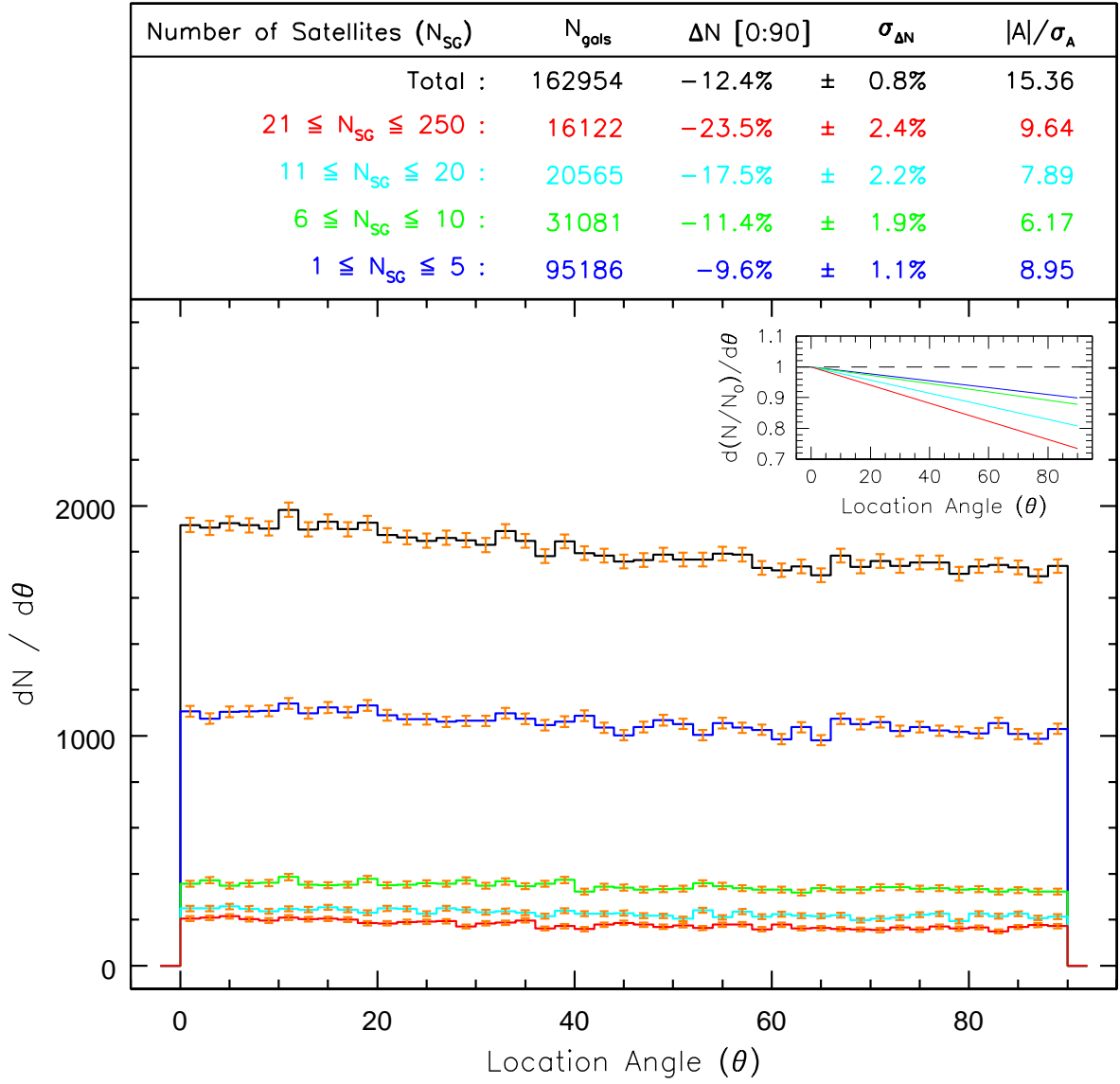}
\caption{\label{f:d1_scount} Satellite location angle ($\theta$) for different
galaxy group sizes (satellites per system, $N_{SG}$). The format of this figure
is identical to that of Figure \ref{f:d1_hg_fdev}. Most satellites exist in
small systems ($N_{SG}\le5$). Alignment strength (fractional slope) increases
with increasing group size. The subsample with the largest groups ($21\le
N_{SG}\le1000$) has the highest trend significance despite having the fewest
member objects, suggesting that major axis preference is more prevalent in
larger galaxy groups. } 
\end{figure}

\clearpage


\begin{figure}
%
\plotone{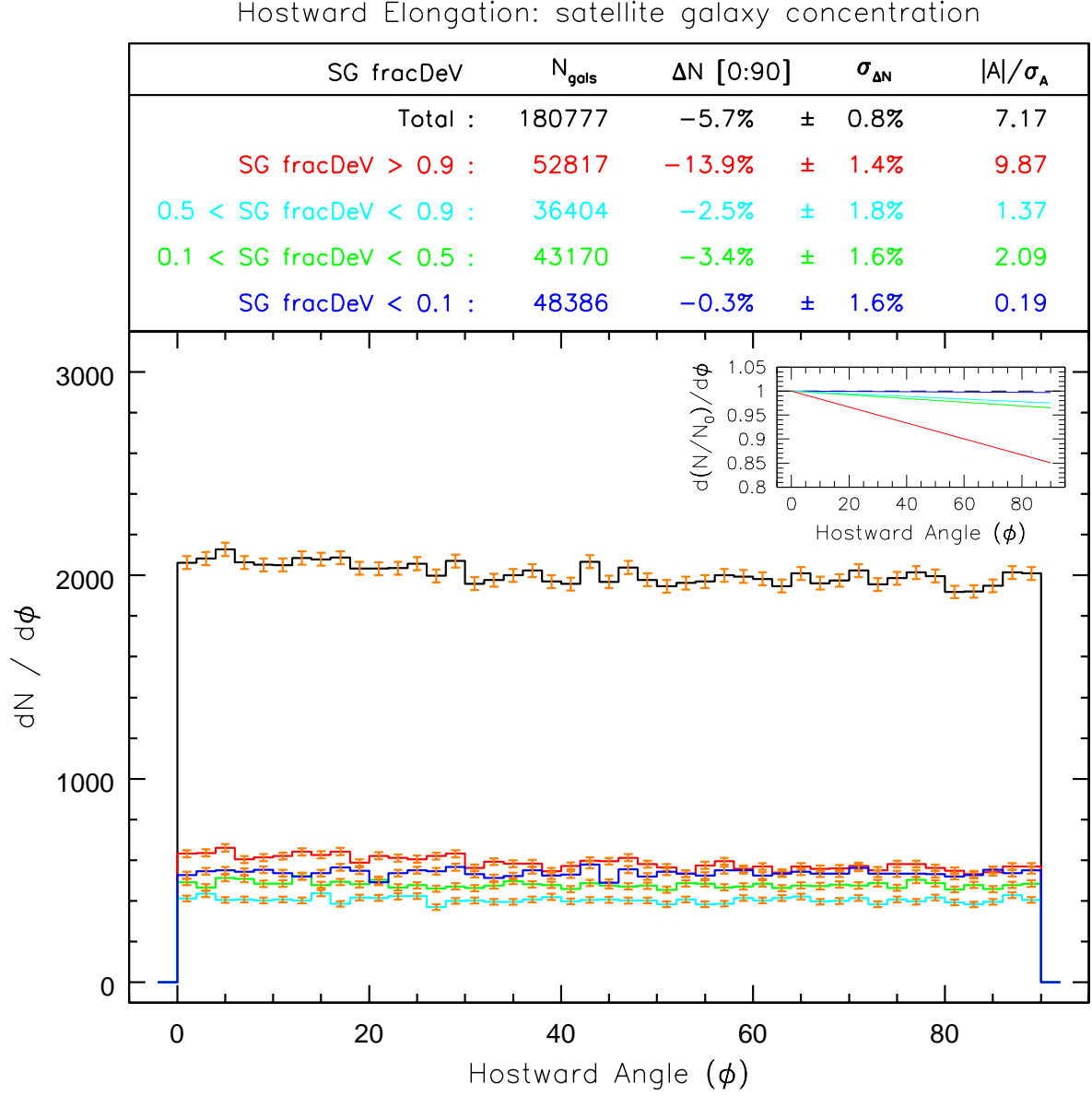}
\caption{\label{f:d3_sg_fdev} Hostward elongation ($\phi$) for different
subsamples of satellite galaxy $\FD$ (concentration). The format of this figure
is identical to that of Figure \ref{f:d1_hg_fdev}. The observed hostward
elongation trend is restricted to high-$\FD$ ($>0.9$) satellite galaxies. }
\end{figure}

\begin{figure}
%
\plotone{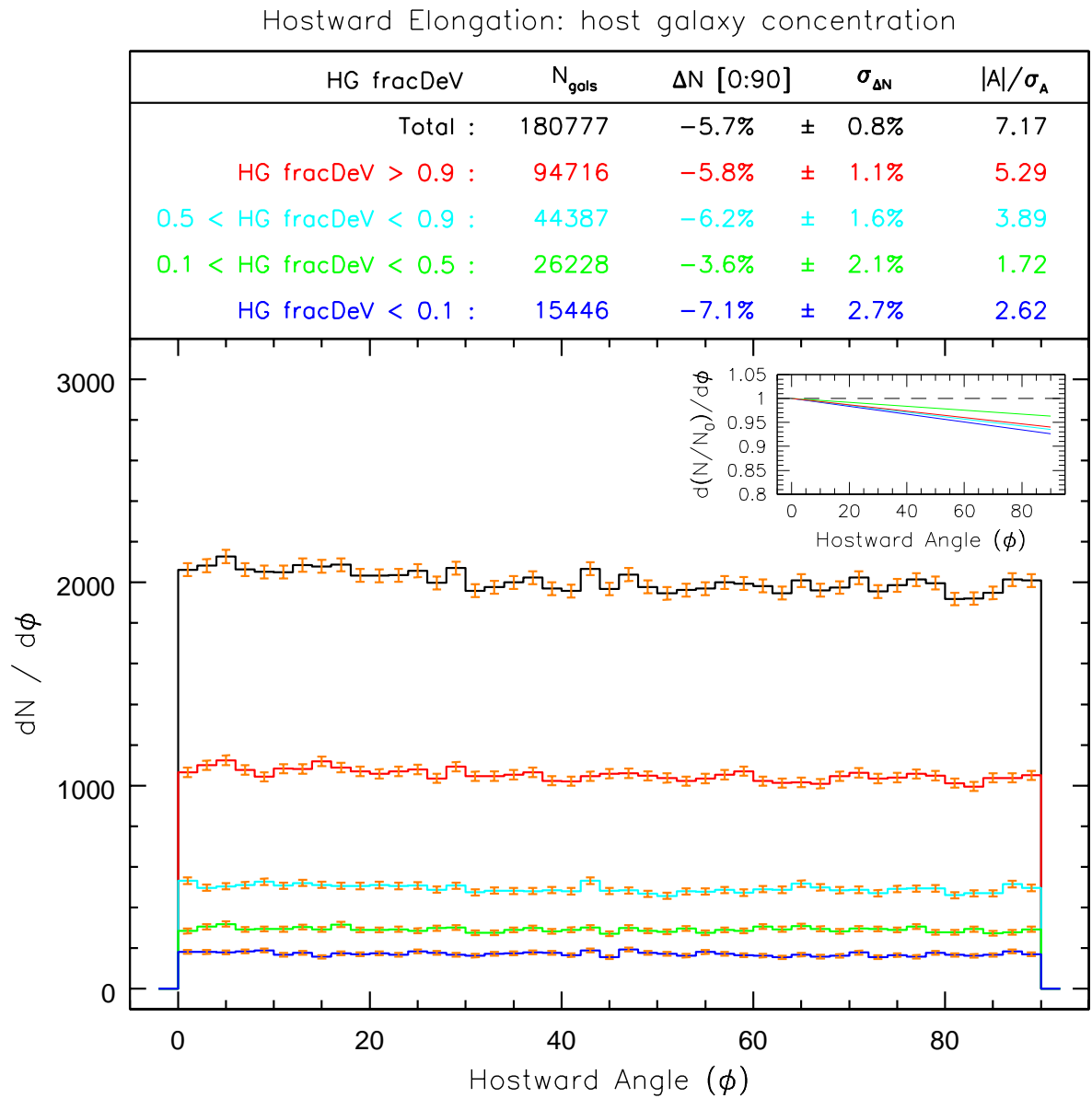}
\caption{\label{f:d3_fdev} Hostward elongation ($\phi$) for different
subsamples of host galaxy $\FD$ (concentration). The format of this figure is
identical to that of Figure \ref{f:d1_hg_fdev}. All four subsamples are
statistically consistent with each other and with the total group sample.
Hostward alignment strength thus appears to be independent of host galaxy
concentration. }
\end{figure}

\begin{figure}
%
\plotone{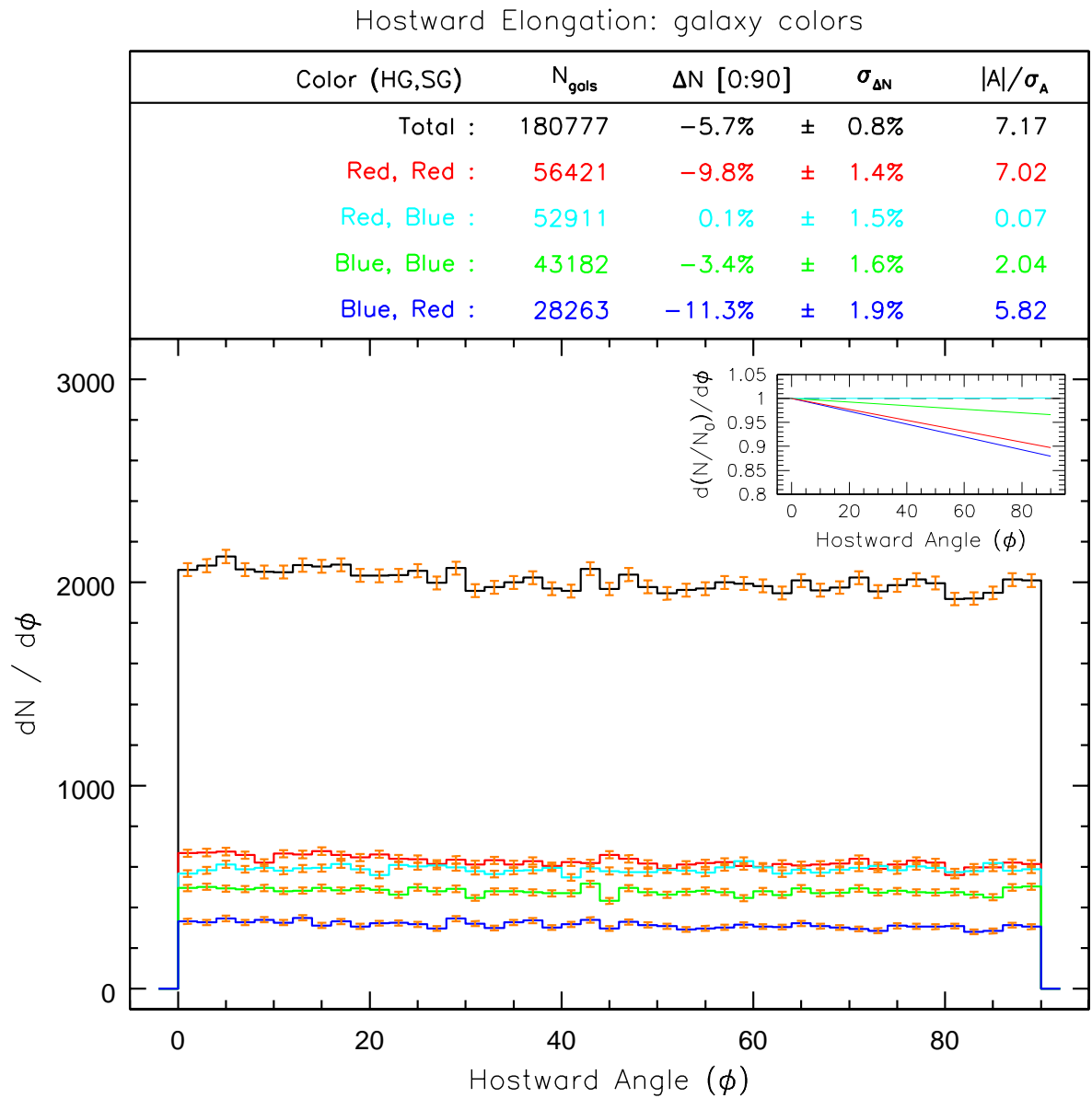}
\caption{\label{f:d3_color} Hostward elongation ($\phi$) as a function of host
and satellite galaxy colors. The format of this figure is identical to that of
Figure \ref{f:d1_hg_fdev}. We perform color separation using the division of
\citet{bailin_2007} (Fig.\ \ref{f:galCMD}). Hostward elongation is restricted
to the two subsamples with red satellite galaxies. Host galaxy color has no
discernible effect on hostward elongation trend strength. } 
\end{figure}

\clearpage

\begin{figure}
%
\plotone{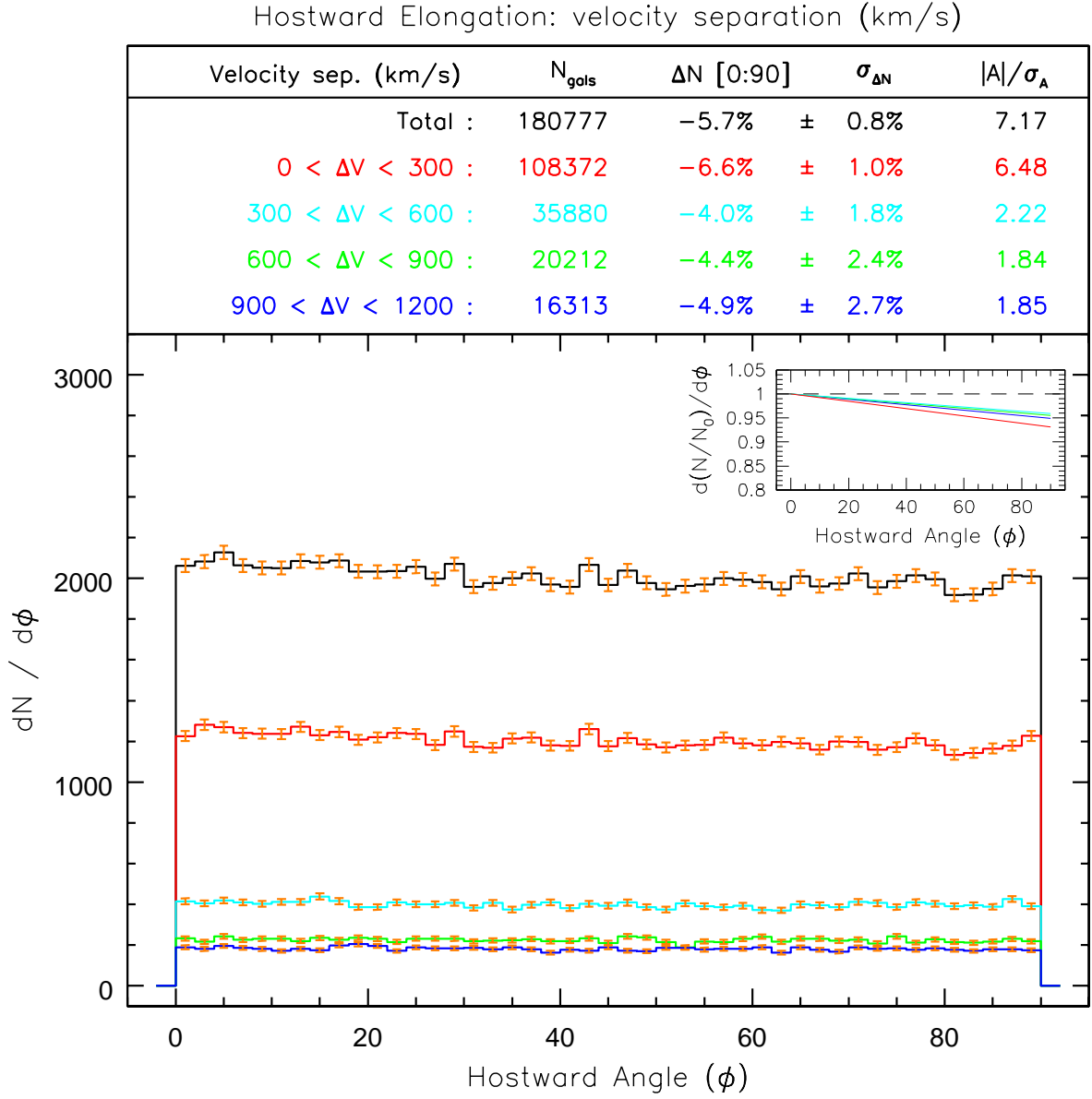}
\caption{\label{f:d3_deltaV} Satellite galaxy hostward elongation ($\phi$) as a
function of velocity separation ($\DV$). The format of this figure is identical
to that of Figure \ref{f:d1_hg_fdev}. Each $\DV$ subsample is statistically
consistent with the total group sample. The subsamples are also loosely
consistent with each other. The high significance in the lowest velocity bin
($|\DV|<300\kms$) is a result of subsample size. On its own, host-satellite
$\DV$ does not influence the degree of hostward alignment. }
\end{figure} 

\begin{figure}
%
\plotone{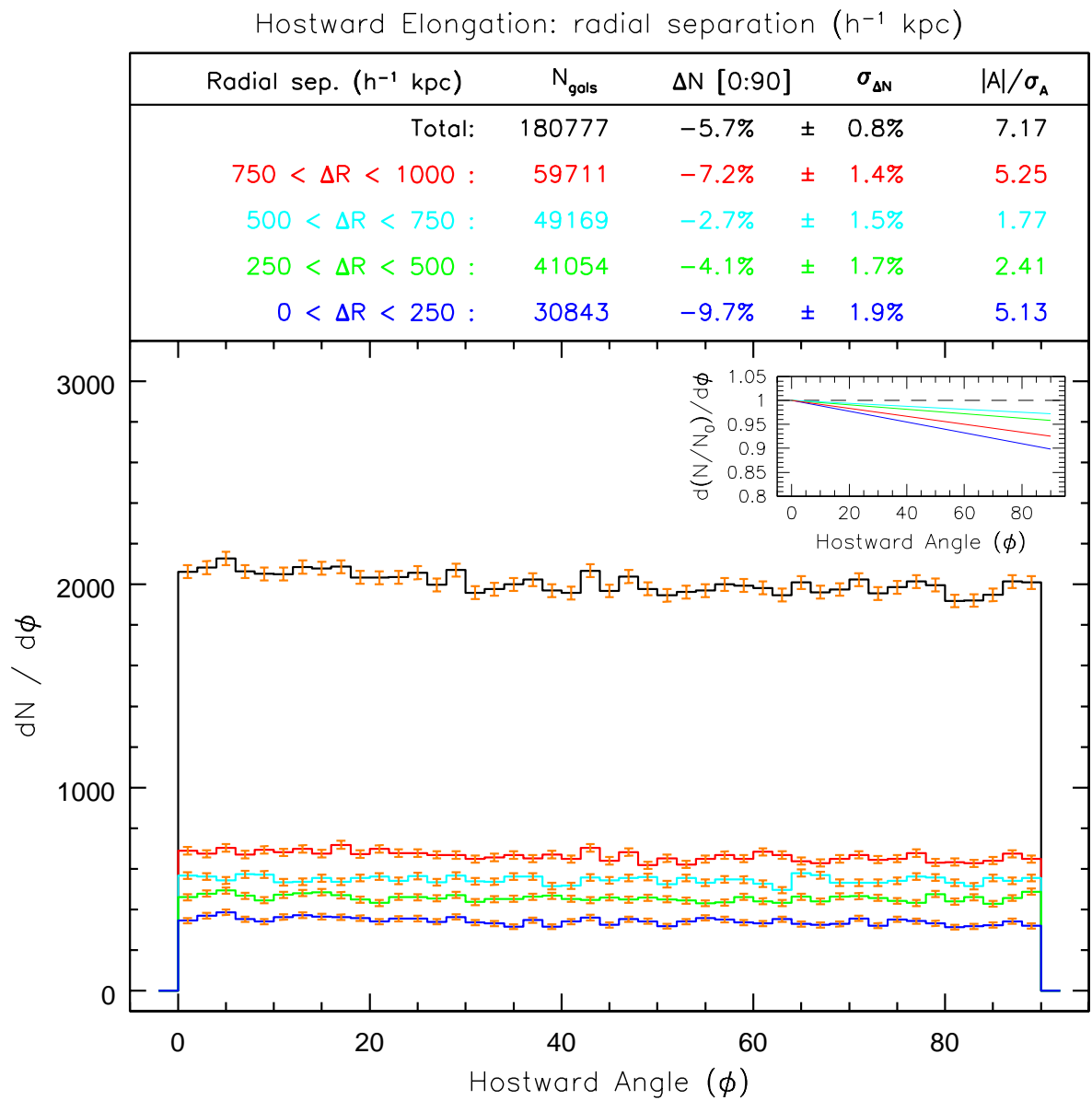}
\caption{\label{f:d3_radius} Satellite galaxy hostward elongation ($\phi$) as a
function of projected radial separation ($\DR$). The format of this figure is
identical to that of Figure \ref{f:d1_hg_fdev}. No subsamples show anisotropy
at very high significance. Hostward alignment is strongest in the innermost
($\DR<250\hkpc$) radial bin. }
\end{figure}

\begin{figure}
%
\plotone{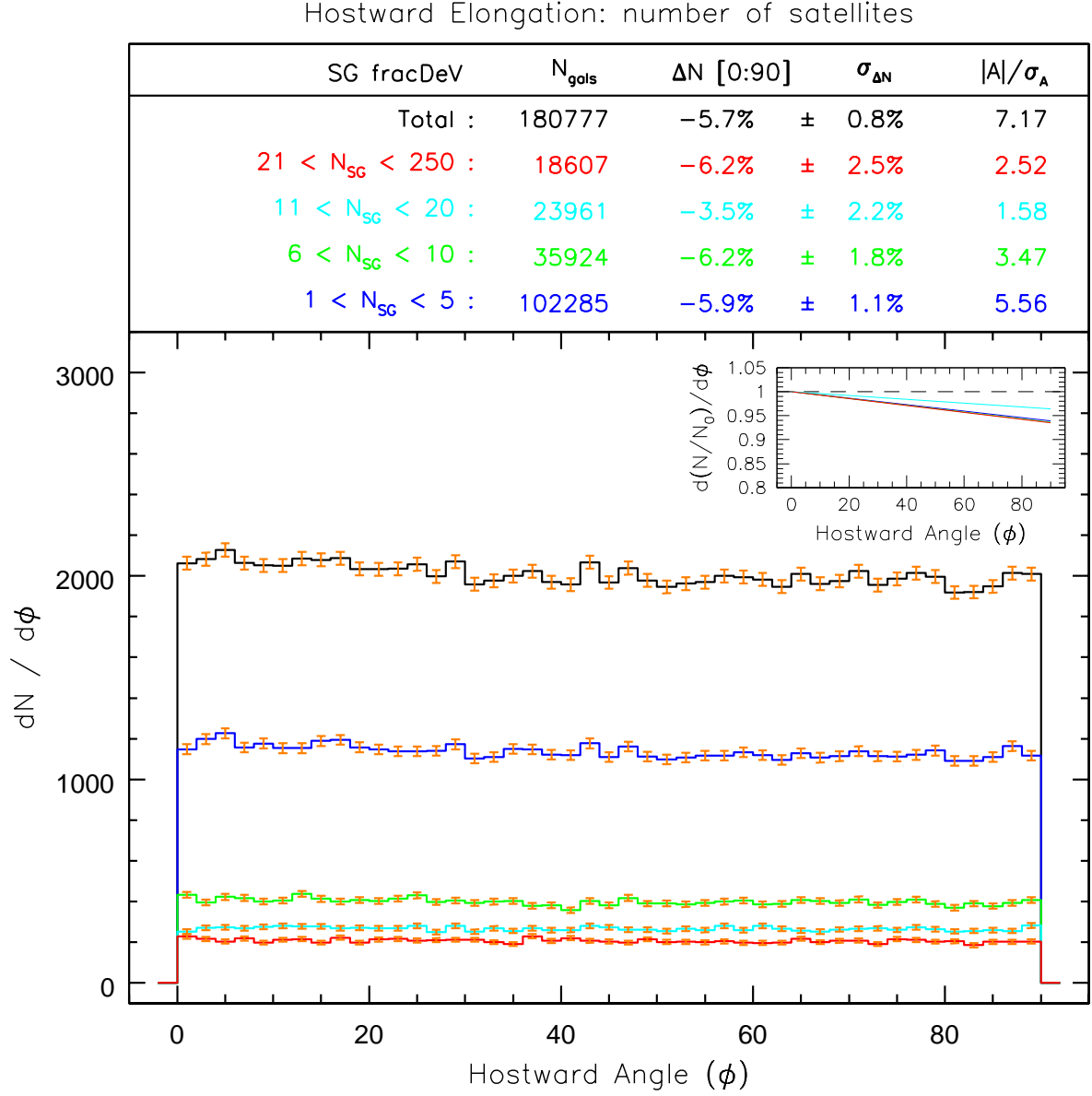}
\caption{\label{f:d3_scount} Satellite location angle as a function of the
number of satellites per host (i.e., group size). The format of this figure is
identical to that of Figure \ref{f:d1_hg_fdev}. All four group size subsamples
are consistent with other and with the total group population. There is no
obvious dependence of hostward elongation on group size. }
\end{figure}

\clearpage

\clearpage

\begin{thebibliography}{}
\label{References}

\bibitem[Abazajian et al.(2004)]{DR2} Abazajian, K., Adelman-McCarthy, J.,
   Agueros, M.A., et al. 2004, \aj, 128, 502

\bibitem[Abazajian et al.(2005)]{DR3} Abazajian, K., Adelman-McCarthy, J., 
   Agueros, M.A., et al. 2005, \aj, 129, 1755


\bibitem[Adelman-McCarthy et al.(2006)]{DR4} Adelman-McCarthy, J., 
   Agueros, M.A., Allam, S.S., et al. 2006, ApJS, 162, 38

\bibitem[Adelman-McCarthy et al.(2008)]{DR6} Adelman-McCarthy, J., 
   Agueros, M.A., Allam, S.S., et al. 2008, ApJS, 175, 297

\bibitem[Agustsson \& Brainerd(2006)]{AB_2006a} Agustsson, I. \& 
    Brainerd, T. G. 2006, \apjl, 644, L25


\bibitem[Argyres et al.(1986)]{ArgEtAl_1986} 
    Argyres, P. C., Groth, E. G., Peebles, P. J. E., \& Struble, M. F. 
    1986, \aj, 91, 471

\bibitem[Assef et al.(2008)]{roberto} 
    Assef, R.J. et al.\ 2008, \apj, 676, 286

\bibitem[Austin \& Peach(1974)]{austin_peach_74} Austin, T. B. \& 
    Peach, J. V. 1974, \mnras, 168, 591

\bibitem[Azzaro et al.(2007)]{azzaro_2007} Azzaro, M., Patiri, S. G., 
    Prada, F., \&  Zentner, A. R. 2007, \mnras, 376, L43



\bibitem[Bailin et al.(2007)]{bailin_2007} Bailin et al.\ 2007, \mnras, 
    submitted (arXiv:0706.1350) 

\bibitem[Bailin et al.(2008)]{bailin_2008} Bailin, J., Power, C., Norberg, P.,
   Zaritsky, D., \& Gibson, B. K. 2008, \mnras, 390, 1133

\bibitem[Bernstein \& Norberg (2002)]{bernberg_2002} Bernstein, G. M. \&
   Norberg, P. 2002, \aj, 124, 733

\bibitem[Bernstein \& Jarvis(2002)]{bernjar} Bernstein, G. M., \& 
    Jarvis, M. 2002, \aj, 123, 583

\bibitem[Binggeli(1982)]{binggeli_1982} Binggeli, B. 1982, \aap, 107, 338


\bibitem[Blanton et al.(2001)]{blant01} Blanton, M. et al. 2001, 
   \aj, 121, 2358


\bibitem[Brainerd(2005)]{brainerd_2005_ani} Brainerd, T. G.  2005,
    \apjl, 628, L101

\bibitem[Carter \& Metcalfe(1980)]{carter_metcalfe_1980} Carter, D. 
    \& Metcalfe, N.  1980, \mnras, 191, 325

\bibitem[Colless et al.(2001)]{2dFGRS} Colless, M. et al. 2001,
   \mnras, 328, 1039

\bibitem[Croft \& Metzler(2000)]{croft_metzler_2000} Croft, R. A. C. \&
   Metzler, C. A. 2000, \apj, 545, 561

\bibitem[Dressler(1978)]{dressler_78} Dressler, A. 1978, \apj, 226, 55

\bibitem[Faltenbacher et al.(2007)]{falten1} Faltenbacher, A. et al., 2007,
    \apjl, 662, L71

\bibitem[Faltenbacher et al.(2008)]{falten2} Faltenbacher, A. et al., 2008,
    \apj, 675, 146

\bibitem[Faltenbacher et al.(2009)]{falten3} Faltenbacher, A., Li, C.,
   White, S. D. M., Jing, Y. P., Mao, S., \& Wang, J. 2009, ReA\&A, 9, 41


\bibitem[Fukugita et al.(1996)]{SDSS_Filters} Fukugita, M., Ichikawa, T., 
    Gunn, J. E., Doi, M., Shimasaku, K., and Schneider, D. P. 
    1996, AJ, 111, 1748

\bibitem[Gunn et al.(1998)]{SDSS_Camera}
    Gunn, J. E., Carr, M. A., Rockosi, C. M., Sekiguchi, M., et al. 
    1998, AJ, 116, 3040

\bibitem[Gunn et al.(2006)]{SDSS_Telescope} 
    Gunn, J. E., Siegmund, W. A., Mannery, E. J., Owen, R. E., et al. 
    2006, AJ, 131

\bibitem[Hirata \& Seljak(2003)]{hirsel} Hirata, C. \& Seljak, U.
   2003, \mnras, 343, 459 

\bibitem[Hirata et al.(2004)]{hirata_2004} Hirata, C. M. et al.
   2004, \mnras, 353, 529

\bibitem[Hogg et al.(2001)]{SDSS_PhotoMonitoring} Hogg, D. W., 
    Finkbeiner, D. P., Schlegel, D. J., and Gunn, J. E. 2001, AJ, 122, 2129

\bibitem[Holmberg(1969)]{holmberg_69} Holmberg, E.  1969, 
    Arkiv for Astronomi, 5, 305

\bibitem[Ivezic et al.(2004)]{SDSS_PhotoQA} Ivezic, Z., Lupton, R. H., 
    Schlegel, D., et al. 2004, AN, 325, 583

\bibitem[Knebe et al.(2004)]{knebe_2004} Knebe, A., Gill, S. P. D., 
   Gibson, B. K., Lewis, G. F., Ibata, R. A. \& Dopita, M. A. 
   2004, \apj, 603, 7




\bibitem[Lee \& Pen(2001)]{lee_pen_2001} Lee, J. \& Pen, U. 
   2001, \apj, 555, 106

\bibitem[Libeskind et al.(2005)]{libesk_2005} Libeskind, N. I., Frenk, C. S., 
   Cole, S., Helly, J. C., Jenkins, A., Navarro, J. F. \& Power, C.
   2005, \mnras, 363, 146 

\bibitem[Libeskind et al.(2007)]{libesk_2007} Libeskind, N. I., Cole, S., 
   Frenk, C. S., Okamoto, T., and Jenkins, A. 2007, \mnras, 374, 16

\bibitem[Lupton et al.(2001)]{SDSS_PHOTO} Lupton, R. H., Gunn, J. E.,
   Ivezi\'{c}, Z., Knapp, G. R., \& Kent, S. 2001,
   in ASP Conf. Ser. 238, Astronomical Data Analysis Software and Systems X,
   ed. F. R. Harnden Jr., F. A. Primini, \& H. E. Payne 
   (San Francisco: ASP), 269

\bibitem[MacGillivray et al.(1976)]{macgillivray76} MacGillivray, H. T.,
   Martin, R., Pratt, N. M., Reddish, V. C., Seddon, H., Alexander, 
   L. W. G., Walker, G. S., \& Williams, P.R. 1976, \mnras, 176, 649

\bibitem[Mandelbaum et al.(2005)]{mandelbaum_2005} Mandelbaum, R. 
   et al. 2005, \mnras, 361, 1287

\bibitem[Matthews et al.(1964)]{matthews_64} Matthews, T. A., Morgan, 
    N. W., \& Schmidt, M. 1964, \apj, 140, 35

\bibitem[Noonan(1972)]{noonan72} Noonan, T. W. 1972, \aj, 77, 9

\bibitem[Padmanabhan et al.(2008)]{SDSS_Ubercal} Padmanabhan, N., et al. 
    2008, ApJ, 674, 1217

\bibitem[Pereira \& Kuhn(2005)]{pereira_kuhn_2005} Pereira, M. J. \&
   Kuhn, J.R. 2005, \apjl, 627, L21

\bibitem[Pier et al.(2003)]{SDSS_Astrometry} Pier, J. R., Munn, J. A., 
    Hindsley, R. B., Hennessy, G. S., Kent, S. M., Lupton, R. H., \& 
    Ivezic, Z. 2003, AJ, 125, 1559

\bibitem[Rhee \& Katgert(1987)]{rhee_katgert_1987} 
    Rhee, G. F. R. N. \& Katgert, P. 1987, \aap, 183, 217

\bibitem[Rhee \& Roos(1989)]{rhee_roos_1989} Rhee, G. F. R. N. \& Roos, N.
    1989, \apss, 157, 201

\bibitem[Rood \& Baum(1967)]{rood_baum_67} 
    Rood, H. J. \& Baum, W. A. 1967, \aj, 72, 398

\bibitem[Rood \& Sastry(1972)]{rood_sastry_1972_A2199} Rood, H. J. \& 
    Sastry, G. N., 1972, \aj, 77, 451

\bibitem[Ryden(2004)]{ryden_04} Ryden, B. S. 2004, \apj, 601, 214

\bibitem[Sastry(1968)]{Sastry_68} Sastry, G. N. 1968, \pasp, 80, 252

\bibitem[Shimasaku et al.(2001)]{Shim01} Shimasaku, K. et al. 2001, 
   \aj, 122, 1238

\bibitem[Smith et al.(2002)]{SDSS_ugriz} Smith, J. A., Tucker, D. L., 
   Kent, S. M., et al. 2002, AJ, 123, 2121

\bibitem[Stoughton et al.(2002)]{SDSS_EDR} Stoughton, C., Lupton, R.H., 
   Bernardi, M., et al. 2002, AJ, 123, 485

\bibitem[Strateva et al.(2001)]{Strat01} Strateva, I. et al. 2001,
   \aj, 122, 1861

\bibitem[Strauss et al.(2002)]{SDSS_GalSamp} 
   Strauss, M.A., Weinberg, D.H., Lupton, R.H. et al. 2002, AJ, 124, 1810

\bibitem[Struble(1987)]{struble_1987} Struble, M. F. 1987, \apj, 317, 668

\bibitem[Struble(1988)]{struble_1988} Struble, M. F. 1988, \aj, 96, 1534

\bibitem[Struble(1990)]{struble_1990} Struble, M. F. 1990, \aj, 99, 743

\bibitem[Struble \& Peebles(1985)]{SP1985}
    Struble, M. F. \& Peebles, P. J. E. 1985, \aj, 90, 582

\bibitem[Trevese et al.(1992)]{trevese92} Trevese, D. et al., 1992, 
    \aj, 104, 935

\bibitem[Tucker et al.(2006)]{SDSS_MTCal} Tucker, D., Kent, S., 
    Richmond, M. W., et al. 2006, AN, 327, 821 

\bibitem[Tucker \& Peterson(1988)]{tucker_peterson_1988} Tucker, G. S. \&
    Peterson, J. B. 1988, \aj, 95, 298

\bibitem[Ulmer et al.(1989)]{ulmer_1989} Ulmer, M.P., McMillan, S.L.W., 
    \& Kowalski, M.P. 1989, \apj, 338, 711

\bibitem[West(1989a)]{west_1989a} West, M.J. 1989, \apj, 344, 535

\bibitem[West(1989b)]{west_1989b} West, M.J. 1989, \apj, 347, 610


\bibitem[Yang et al.(2006)]{yang_2006} Yang, X., van den Bosch, F. C., 
    Mo, H. J., Mao, S., Kang, X., Weinmann, S. M., Guo, Y. \& Jing, Y. P.
    2006, \mnras, 369, 1293

\bibitem[Yasuda et al.(2001)]{yasuda01} Yasuda, N. et al. 2001, \aj, 122, 1104

\bibitem[York et al.(2000)]{york2000} 
    York, D. G., Adelman, J., Anderson, J. E., et al. 
    2000, AJ, 120, 1579

\bibitem[Zaritsky et al.(1997a)]{zaritsky_97_catalog} 
    Zaritsky, D., Smith, R., Frenk, C. S., \& White, S. D. M. 1997a, 
    \apj, 478, 39

\bibitem[Zaritsky et al.(1997b)]{zaritsky_97_ani} 
    Zaritsky, D., Smith, R., Frenk, C. S., \& White, S. D. M. 1997b, 
    \apjl, 478, L53

\bibitem[Zentner et al.(2005)]{zentner_2005} Zentner, A. R., Kravtsov, A. V.,
   Gnedin, O. Y., \& Klypin, A. A. 2005, \apj, 629, 219

\end{thebibliography}
\end{document}